\documentclass{article}
\usepackage[utf8]{inputenc}
\usepackage[english]{babel}
\usepackage{amsmath}
\usepackage{amssymb}
\usepackage{amsthm}
\usepackage{tikz}
\usetikzlibrary{petri}
\usepackage[matrix,arrow,curve]{xy}
\usepackage{hyperref}
\usepackage{mathpartir}
\usepackage{breakurl}
\usepackage{wrapfig}
\usepackage{longtable}

\newcommand{\Section}[1]{Section~\ref{#1}}

\newcommand{\longtwoheadrightarrow}{\longrightarrow\hspace{-1.2em}\rightarrow\hspace{.2em}}
\newcommand{\transitionpath}[1]{\stackrel{#1}{\longtwoheadrightarrow}}
\newcommand{\Id}{\mathop{\mathrm{Id}}}

\newcommand{\vxym}[1]{\vcenter{\xymatrix{#1}}}
\newcommand{\svxym}[1]{\vcenter{\xymatrix@R=3ex@C=3ex{#1}}}

\newcommand{\N}{\mathbb{N}}
\newcommand{\slicecat}[2]{#1\mathop{\downarrow}#2}
\newcommand{\category}[1]{\mathbf{#1}}
\newcommand{\Set}{\category{Set}}

\newcommand{\CSet}{\category{CSet}}
\newcommand{\nCSet}[1]{\category{CSet}_{#1}}
\newcommand{\PCSet}{\category{PCSet}}
\newcommand{\nPCSet}[1]{\category{PCSet}_{#1}}
\newcommand{\SCSet}{\category{SCSet}}
\newcommand{\nSCSet}[1]{\category{SCSet}_{#1}}
\newcommand{\SPCSet}{\category{SPCSet}}

\newcommand{\LSCSet}{\category{LSCSet}}
\newcommand{\nLSCSet}[1]{\category{LSCSet}_{#1}}
\newcommand{\wLSPCSet}{\category{wLSPCSet}}
\newcommand{\LSPCSet}{\category{LSPCSet}}

\newcommand{\HDA}{\category{HDA}}
\newcommand{\nHDA}[1]{\category{HDA}_{#1}}
\newcommand{\sHDA}{\category{sHDA}}
\newcommand{\nsHDA}[1]{\category{sHDA}_{#1}}

\newcommand{\PNet}{\category{PNet}}
\newcommand{\sPNet}{\category{sPNet}}
\newcommand{\LPNet}{\category{LPNet}}
\newcommand{\sLPNet}{\category{sLPNet}}

\newcommand{\ccat}{\square}
\newcommand{\sccat}{\square_S}
\newcommand{\nsccat}[1]{(\square_S)_{#1}}
\newcommand{\pccat}{\boxdot}
\newcommand{\spccat}{\boxdot_S}
\newcommand{\Pre}{\mathop{\mathrm{pre}}}
\newcommand{\Post}{\mathop{\mathrm{post}}}
\newcommand{\pre}[1]{\null^\bullet #1}
\newcommand{\post}[1]{#1^\bullet}
\newcommand{\transition}[1]{\stackrel{#1}{\longrightarrow}}
\newcommand{\id}{\mathrm{id}}
\newcommand{\op}{\mathrm{op}}
\newcommand{\qeq}{\quad=\quad}
\newcommand{\qcomma}{,\quad}

\newcommand{\qqtand}{\qquad\text{and}\qquad}
\newcommand{\resp}{resp.~}
\newcommand{\nbd}{\nobreakdash-\hspace{0pt}}

\newcommand{\setof}[1]{\{\ #1\ \}}
\newcommand{\tq}{\ |\ }
\newcommand{\hda}{\mathrm{hda}}
\newcommand{\pn}{\mathrm{pn}}
\newcommand{\lbl}[2]{\ar@{}[#1]|-{#2}}

\newcommand{\lcs}[1]{\mathop{!}#1} 
\newcommand{\slcs}[1]{\lcs{#1}}
\newcommand{\ie}{i.e.~}
\newcommand{\wrt}{wrt~}
\renewcommand{\geq}{\geqslant}
\renewcommand{\leq}{\leqslant}
\newcommand{\fpointed}{\mathop{?}}
\newcommand{\pPCSet}{\category{PCSet}_{\fpointed}}
\newcommand{\pointed}[1]{#1^*}
\newcommand{\ptSet}{\pointed\Set}

\newcommand{\CTS}{\category{CTS}}

\newcommand{\TS}{\category{TS}}
\newcommand{\sTS}{\category{sTS}}

\newcommand{\indep}[1]{\mathop{I_{#1}}}
\newcommand{\indepp}[1]{\mathop{I'_{#1}}}
\newcommand{\ACR}{\category{ACR}}
 \newcommand{\sACR}{\category{sACR}}

\newcommand{\incompat}{\mathop{\#}}
\newcommand{\ES}{\category{ES}}
\newcommand{\sES}{\category{sES}}
\newcommand{\LES}{\category{LES}}
\newcommand{\sLES}{\category{sLES}}

\hypersetup{
  pdftitle={Formal Relationships Between Geometrical and Classical Models for Concurrency},
  pdfauthor={Éric Goubault and Samuel Mimram},
  unicode=true,
  colorlinks=true,
  linkcolor=black,
  citecolor=black,
  urlcolor=black
}

\newtheorem{theorem}{Theorem}
\newtheorem{definition}[theorem]{Definition}

\newtheorem{proposition}[theorem]{Proposition}
\newtheorem{lemma}[theorem]{Lemma}
\theoremstyle{remark}
\newtheorem{remark}[theorem]{Remark}
\theoremstyle{example}
\newtheorem{example}[theorem]{Example}

\renewcommand{\C}{\mathcal{C}}
\newcommand{\D}{\mathcal{D}}

\renewcommand{\paragraph}[1]{\medskip\noindent\textbf{#1}}

\title{Formal Relationships Between Geometrical and Classical Models for
  Concurrency}

\author{Éric Goubault and Samuel Mimram\thanks{CEA, LIST, Point Courrier 94,
    91191 Gif-sur-Yvette, France. This work has been supported by the PANDA
    (``Parallel and Distributed Analysis'', \hbox{ANR-09-BLAN-0169}) French ANR
    project.}}

\begin{document}
\maketitle

  \begin{abstract}
    A wide variety of models for concurrent programs has been proposed during
    the past decades, each one focusing on various aspects of computations:
    trace equivalence, causality between events, conflicts and schedules due to
    resource accesses, etc. More recently, models with a geometrical flavor have
    been introduced, based on the notion of cubical set. These models are very
    rich and expressive since they can represent commutation between any number
    of events, thus generalizing the principle of \emph{true concurrency}. While
    they are emerging as a central tool in concurrency, which is very promising
    because they make possible the use of techniques from algebraic topology in
    order to study concurrent computations, they have not yet been precisely
    related to the previous models, and the purpose of this paper is to fill
    this gap. In particular, we describe an adjunction between Petri nets and
    cubical sets which extends the previously known adjunction between Petri
    nets and asynchronous transition systems by Nielsen and
    Winskel.
  \end{abstract}











A great variety of models for concurrency was introduced in the last decades:
transition systems (with independence), asynchronous automata, event structures,
Petri nets, etc. Each of these models focuses on modeling a particular aspect of
computations, and even though their nature are very different, they are tightly
related to each other as witnessed in~\cite{winskel-nielsen:models-concur}. More
recently, models inspired by ideas coming from geometry, such as \emph{cubical
  sets} (also sometimes called \emph{higher dimensional automata} or
HDA~\cite{VPModeling,EGTPJHomology}) or local po-spaces~\cite{LFEGMRAlgebraic},
have emerged as central tools to study concurrency: thanks to their nice
algebraic structure, they allow one to carry on abstractly many computations,
and they are very expressive because of their ability to represent commutations
between multiple events. However, since their introduction, they have not been
systematically and formally linked with the other models, such as transition
systems, even though cubical sets contain a notion of generalized transition in
their very definition.

From a scientific point of view, the mere observation that these models are
different is not satisfactory and their links with other models have to be
investigated in depth. However, it turns out that their relationship is often
quite subtle: the various models are usually not isomorphic, nor even one is a
retract of the other. Adjunctions between the categories of models, which
generalize Galois connections to categories, are the right notion to relate and
compare them. This was first studied in the context of operational models for
concurrency by Winskel et al.~\cite{winskel-nielsen:models-concur} and extended
to geometrical models~\cite{EGGeometry}, but only between fairly restricted
categories. In this paper, we greatly improve previous work by extending it to
the full categories of transition systems (operational model of ``interleaving''
concurrency) and of transition systems with independence (operational model of
``true'' concurrency). Another approach to compare these models, based on
history-preserving bisimulations, is developed
in~\cite{van2006expressiveness}. The main motivation underlying this work is
that, by relating these models, we can compare the semantics of concurrent
languages given in different formalisms.
This also allows for reusing specific methods for statically analyzing
concurrent programs in one model (such as deadlock detection algorithms for
cubical sets~\cite{LFEGMRDetecting}, invariant generation on Petri
nets~\cite{invPetri}, state-space reduction techniques such as sleep sets and
persistent sets in Mazurkiewicz traces~\cite{PGPWUsing}, or stubborn sets in
Petri nets~\cite{AVStubborn2}) in the other.

This paper constitutes a major step towards formally relating geometric models
with other models for concurrency. The links might appear as intuitive, but the
formal step we are making underlines subtle differences between the models:
there are many variants of the models, all of which can be embedded in the model
of HDA, which allows us to precisely characterize the outcomes of choosing one
of the other variant of the models. We have done our best to express in
categorical terms how to construct one variant from the other. In particular,
most models admits the following variations:
\begin{itemize}
\item events can be labeled or not,
\item morphisms can be strict or partial,
\item the multiplicity of an event can be taken in account or not,
\item in the case where the events are labeled, morphisms between labels can be
  strict or not.
\end{itemize}
It turns out from this study that \emph{strongly labeled} HDA seem to be the
right notion of HDA, at least for comparing with most other common models of
concurrency.  This also unravels interesting phenomena (besides being necessary
for being able to relate semantics given in different styles)
such as the fact that persistent set types of methods for tackling the
state-space explosion problem can be seen as searching for retracts of the state
space, in the algebraic topological sense.
We end this article by making some hypotheses on further relationships, with
event structures and Petri nets in particular.

\paragraph{Related work.}
In this paper, we extend Winskel's results~\cite{winskel-nielsen:models-concur},
which include adjunctions between transition systems, event structures, trace
languages, asynchronous transition systems and Petri nets which are still an
active research area~\cite{staton2010expressivity}. A first step towards
comparing higher-dimensional automata (a form of geometric semantics we are
considering here), Petri nets, and event structures is reported
in~\cite{vanglabbeek:pn-cs-hda}. Also, an investigation of the comparison
between cubical sets (another form of geometric semantics) and transition
systems, as well as transition systems with independence was started
in~\cite{EGCubical}, but never formally published.

We describe right adjoint functors from the categories of transition systems,
asynchronous transition systems, Petri nets and prime event structures
of~\cite{winskel-nielsen:models-concur}, to HDA. By general theorems, these
functors transport limits onto limits, hence preserve classical parallel
semantics based on pullbacks, by synchronized products~\cite{AASystemes}, as the
ones in transition systems or the ones of
\cite{winskel-nielsen:models-concur}.
Cubical sets (or more generally HDA) that we take as the primary model for
geometric semantics here, have appeared in numerous previous works, in algebraic
topology in particular~\cite{JPSHomologie,RBPJHAlgebra}.  A monoidal
presentation
can also be found in \cite{grandis-mauri:cubical-sets}. The basics of ``directed
algebraic topology'' that is at the basis of the mathematics involved in the
geometric semantics we use here can be found in~\cite{Grandisbook}.

\paragraph{Contents  of the paper.}
We begin by recalling the geometric model provided by cubical sets
in \Section{sec:geometric-models} and some well-known models for concurrent computations
(transition systems, asynchronous automata, event structures and Petri nets) in
\Section{sec:traditional-models}. We then relate them by defining adjunctions
in \Section{sec:adjunctions}. HDA naturally ``contain'' transition systems
(resp. asynchronous transition systems), which just encode the non-deterministic
(resp. and pairwise independence) information. Event structures are also shown
to be more abstract than HDA: they impose binary conflict relations and
conjunctive dependencies (an event cannot depend on a disjunction of two
events), and they do not distinguish different occurrences of the same
event. Petri nets have a built-in notion of degree of parallelism, as is the
case of HDA (given by cell dimension) but impose specific constraints on
dynamics.
We
finally conclude on future works in \Section{sec:conclusion}.

\section{Geometric models for concurrency}
\label{sec:geometric-models}
Precubical sets can be thought as some sort of generalized transition systems
with higher-dimensional transitions. Similarly to transition systems there is a
corresponding notion with ``idle transitions'', called \emph{cubical
  sets}. These classical objects in combinatorial algebraic topology (see for
instance \cite{JPSHomologie}) have been used as an alternative \emph{truly
  concurrent} model for concurrency, in particular since the seminal papers
\cite{VPModeling} and \cite{RVGBisimulation}. More recently, they have been used
in \cite{LFEGMRDetecting} and \cite{LFEGMRAlgebraic} for deriving new and
interesting deadlock detection algorithms. More algorithms have been designed
since then, see for instance \cite{MRMSCS} and \cite{LFMSCS}.
In the following, we will be mostly using \emph{symmetric precubical
  sets}. However, we have done our best to introduce here the notion gradually,
and recall some variants as well as important properties.

\subsection{Cubical sets}
\label{sec:cset}

A cubical set consists of a family $(C(n))_{n\in\N}$ of sets, the elements
of~$C(n)$ being called \emph{$n$-cells}, together with for every pairs of
integers $n$ and $i$, such that $0\leq i\leq n$, maps
\[
\partial_i^-,\partial_i^+:C(n+1)\to C(n)
\qqtand
\iota_i:C(n)\to C(n+1)
\]
respectively called \emph{source}, \emph{target} and \emph{degeneracy maps},
satisfying
\begin{equation}
  \label{eq:cset1}
  \partial_j^\beta\partial_i^\alpha=\partial_i^\alpha\partial_{j-1}^\beta
  \qquad\qquad
  \iota_i\iota_j=\iota_{j-1}\iota_i
\end{equation}
with $i<j$ and~$\alpha,\beta\in\{-,+\}$ and, for every~$\alpha\in\{-,+\}$,
\begin{equation}
  \label{eq:cset2}
  \partial_j^\alpha\iota_i=
  \begin{cases}
    \iota_i\partial_{j-1}^\alpha
    &\text{if~$i<j$}\\
    \id &\text{if~$i=j$}\\
    \iota_{i-1}\partial_j^\alpha
    &\text{if~$i>j$}\\
  \end{cases}
\end{equation}
A morphism~$\kappa:C\to C'$ between two cubical sets~$C$ and~$C'$ consists of a
family $(\kappa_n:C(n)\to C'(n))_{n\in\N}$ of functions which is natural: for
every index~$i$ and~$\alpha\in\{-,+\}$,
\[
\kappa_n\circ\partial_i^\alpha=\partial_i^\alpha\circ\kappa_{n+1}
\qqtand
\kappa_{n+1}\circ\iota_i=\iota_i\circ\kappa_n
\]
and we write~$\CSet$ for the category thus defined. The~\emph{0-source} (\resp
\emph{0-target}) of an $n$-cell $x\in C(n)$ is the $0$-cell
$\partial_0^-\ldots\partial_0^-(x)$ (\resp $\partial_0^+\ldots\partial_0^+(x)$).

More conceptually, a \emph{cubical set}~$C$ is a presheaf on the cubical
category~$\ccat{}$, that is a functor \hbox{$C:\ccat{}^\op\to\Set$}, and a
morphism of cubical sets is a natural transformation between the corresponding
functors. Here, the \emph{cubical category}~$\ccat{}$ is defined as the free
category on the graph whose objects are natural integers~$n\in\N$ and
containing, for every integers $i$ and~$n$ such that $0\leq i\leq n$ and every
$\alpha\in\{-,+\}$, arrows
\begin{equation}
  \label{eq:ccat-arrows}
  \varepsilon_{i,n}^\alpha:n\to n+1
  \qqtand
  \eta_{i,n}:n+1\to n
\end{equation}
quotiented by the relations expressing axioms dual to those given for cubical
sets~\eqref{eq:cset1} and~\eqref{eq:cset2} -- so that for every index~$n$, the
function $C(\varepsilon_{i,n}^\alpha)$ corresponds to $\partial_i^\alpha$ and
$C(\eta_{i,n})$ corresponds to $\iota_i$:
\begin{equation}
  \label{eq:ccat-eqs}
  \varepsilon_{i,n+1}^\beta\varepsilon_{j,n}^\alpha
  =
  \varepsilon_{j-1,n}^\alpha\varepsilon_{i,n+1}^\beta
  \qquad\qquad
  \eta_{j,n}\eta_{i,n+1}=\eta_{i,n}\eta_{j-1,n+1}
\end{equation}
with~$i<j$ and $\alpha,\beta\in\{-,+\}$, and for every $\alpha\in\{-,+\}$,
\[
\eta_{i,n}\varepsilon_{j,n}^\alpha
\qeq
\begin{cases}
  \varepsilon_{j-1,n-1}^\alpha\eta_{i,n-1}
  &\text{if~$i<j$}\\
  \id&\text{if~$i=j$}\\
  \varepsilon_{j,n-1}^\alpha\eta_{i-1,n-1}
  &\text{if~$i>j$.}
\end{cases}
\]

The \emph{precubical category}~$\pccat$ is defined similarly with only the
$\varepsilon_{i,n}^\alpha$ as generators and the first equations
of~\eqref{eq:ccat-eqs} as axioms, and a \emph{precubical set} is a presheaf on
the precubical category: a precubical set consists of a family $(C(n))_{n\in\N}$
of sets together with a family of maps $\partial_i^-,\partial_i^+:C(n+1)\to
C(n)$ satisfying the equations on the left of~\eqref{eq:cset1}. We
write~$\PCSet$ for the corresponding category.

Given an integer~$n$, we write~$\ccat_n$ for the full subcategory of~$\ccat$
whose objects are the integers $k\leq n$. An \emph{$n$-dimensional cubical set}
is a presheaf on~$\ccat_n$ and we write~$\nCSet{n}$ for the category of
$n$-dimensional cubical sets. The inclusion functor~$\ccat_n\to\ccat$ induces by
precomposition a functor \hbox{$U_n:\CSet\to\nCSet{n}$} called the
\emph{$n$-truncation functor} (see \Section{sec:cs-relate}).

\begin{example}
  The geometric intuition underlying cubical sets is the following one. An
  $n$\nbd{}cell~$x$ of a cubical set should be seen as an $n$-dimensional cube,
  the \hbox{$(n-1)$}\nbd{}dimensional cubes $\partial_i^-(x)$ and
  $\partial_i^+(x)$ being respectively the source and target in dimension~$i$
  of~$x$, and the degeneracy maps~$\iota_i$ allowing us to see an
  $n$-dimensional cube as an $(n+1)$\nbd{}dimensional one, degenerated in
  dimension~$i$. So for example, a ``cylinder'' can be described as a precubical
  set~$C$ with
  \[
  C(0)=\{x,y\}
  \qquad
  C(1)=\{f,g,h\}
  \qquad
  C(2)=\{\alpha\}
  \qquad
  C(n)=\emptyset\quad\text{for $n>2$}
  \]
  with the following sources and targets, given by
  $\partial_0^-(f)=\partial_0^+(f)=\partial_0^-(h)=x$,
  $\partial_0^-(g)=\partial_0^+(g)=\partial_0^+(h)=y$,
  $\partial_0^-(\alpha)=\partial_0^+(\alpha)=h$, $\partial_1^-(\alpha)=f$ and
  $\partial_1^+(\alpha)=g$. This cylinder can be pictured graphically as
  \[
  \vxym{
    &y\ar@(ur,dr)^g\ar@{}[d]|-\alpha\\
    x\ar@(ur,dr)^f\ar[ur]^h&\\
  }
  \qquad\text{or}\qquad
  \vxym{
    x\ar[r]^h\ar[d]_f\ar@{}[dr]|-\alpha&y\ar[d]^g\\
    x\ar[r]_h&y
  }
  \quad\text{(in an unfold representation)}
  \]
  From a concurrency point of view, a 1-cell corresponds to the occurrence of an
  event (an action) and an $n$-cell corresponds to a commutation or an
  independence between the 1-cells occurring in its faces. The cubical set above
  representing the cylinder thus corresponds intuitively to a program
  constituted of two processes in parallel: a (while) loop (the actions $f$ and
  $g$) and a single instruction ($h$). See also Example~\ref{ex:lcs}.
\end{example}

In previous example, the two transitions~$f$ and~$g$ are instances of a same
event because they are parallel faces of the square~$\alpha$. This suggests that
the notion of event should be reconstructed in a precubical set as an
equivalence class of transitions as follows. Suppose given a precubical
set~$C$. We define a relation~$\approx$ as the smallest equivalence relation on
1-cells of $C$, such that for every \hbox{$f,g\in C(1)$}, $f\approx g$ when
there exists~$y\in C(2)$ such that $f=\partial_i^-(y)$ and~$g=\partial_i^+(y)$,
for $i=0$ or $i=1$. An \emph{event} is the equivalence class of a 1-cell under
the relation~$\approx$. Given a morphism~$\kappa:C\to D$ between precubical
sets, two 1-cells of~$D$ in a same event are sent to two 1-cells of~$D$ in a
same event; any such morphism thus induces a function~$\kappa_1/\!\approx$ from
the events of~$C$ to the events of~$D$.

\subsection{A monoidal definition of the cubical category}
\label{sec:mon-cc}
A shorter description of the cubical category can be given if we take its
monoidal structure in account: the cubical category is the free monoidal
category (that is, a category equipped with a coherent tensor product and
unit~\cite{maclane:cwm}) containing a co\nbd{}cubical
object~\cite{grandis-mauri:cubical-sets}. This will help in defining very
concisely the adjunctions we have in mind in Section~\ref{sec:adjunctions}.

\begin{definition}
  A \emph{cubical object} $(C,\varepsilon^-,\varepsilon^+,\eta)$ in a monoidal
  category~$(\C,\otimes,I)$ consists of an object~$C$ together with three
  morphisms
  \[
  \eta:I\to C
  \qquad\qquad
  \varepsilon^-:C\to I
  \qquad\qquad
  \varepsilon^+:C\to I
  \]
  such that
  \[
  \varepsilon^-\circ\eta
  \qeq
  \id_I
  \qeq
  \varepsilon^+\circ\eta
  \]
  A \emph{morphism}~$f$ between two cubical
  objects~$(C_1,\varepsilon^-_1,\varepsilon^+_1,\eta_1)$ and
  $(C_2,\varepsilon^-_2,\varepsilon^+_2,\eta_2)$ is a morphism~$f:C_1\to C_2$
  such that
  \[
  f\circ\eta_1=\eta_2
  \qquad\qquad
  \varepsilon_2^-\circ f=\varepsilon_1^-
  \qquad\qquad
  \varepsilon_2^+\circ f=\varepsilon_1^+
  \]
  Dually, a \emph{co-cubical object} $(C,\varepsilon^-,\varepsilon^+,\eta)$
  in~$\C$ is a cubical object in~$\C^\op$.
\end{definition}

\noindent
In the cubical category~$\ccat$, $(1,\varepsilon^-,\varepsilon^+,\eta)$ is a
co-cubical object.
The fact that~$\ccat$ is the free monoidal category containing a co-cubical
object means that all the arrows of~$\ccat$ can be recovered from those by
tensoring with identities
\[
\varepsilon_{i,n}^\alpha=\id_i\otimes\varepsilon^\alpha\otimes\id_{n-i}
\qqtand
\eta_{i,n}=\id_i\otimes\eta\otimes\id_{n-i}
\]
and that the axioms satisfied by the morphisms -- the axioms dual
of~\eqref{eq:cset1} and~\eqref{eq:cset2} -- are precisely those imposed by the
axioms of monoidal categories and those of co-cubical objects. This can be
equivalently reformulated as follows:

\begin{proposition}
  Given a monoidal category~$\C$, the category of monoidal functors~$\ccat\to\C$
  and monoidal natural transformations is equivalent to the category of
  co-cubical objects in~$\C$.
\end{proposition}

\noindent
In other words, given a monoidal category~$\C$, a cubical object in~$\C$ is
``the same'' as a monoidal functor~$\ccat{}^\op\to\C$. This definition of
cubical sets has been known for quite some time, but no concrete application of
it has been done up to now. Interestingly, we show here that it can be used to
concisely define some cubical sets (see in particular \Section{sec:lcs}). It is
also sometimes useful to define morphisms; for instance, given integers~$n$
and~$i$ such that~$0\leq i\leq n$, and~$\alpha\in\{-,+\}$, we
write~$\partial_{\lnot i}^\alpha:C(n+1)\to C(1)$ for the morphism
$\partial_{\lnot i}^\alpha=C((\varepsilon^\alpha)^{\otimes
  i}\otimes\id_1\otimes(\varepsilon^\alpha)^{\otimes(n-i)})$
where~$(\varepsilon^\alpha)^{\otimes i}$ denotes the tensor product of~$i$
copies of~$\varepsilon^\alpha$.

Similarly, monoidal functors~$\pccat\to\C$ correspond to co-precubical objects
in~$\C$, where a \emph{precubical object} $(C,\varepsilon^-,\varepsilon^+)$ is
an object~$C$ of~$\C$ together with two arrows~$\varepsilon^-,\varepsilon^+:C\to
I$ (and no axiom to be satisfied), and a \emph{co-precubical object} is defined
dually.

\subsection{From precubical sets to cubical sets}
\label{sec:partial-pcset}
In this section, we formalize the intuition that morphisms between precubical
sets are to morphisms between cubical sets what partial functions are to total
functions. Recall that a \emph{pointed set} $(A,a)$ consists of a set together
with a distinguished element $a\in A$, and a morphism $f:(A,a)\to(B,b)$ between
two pointed sets consists of a function~$f:A\to B$ such that~$f(a)=b$. If we
write~$\ptSet$ for the category of pointed sets, there is a forgetful functor
$U:\ptSet\to\Set$ which to every pointed set~$(A,a)$ associates the underlying
set~$A$. This functor admits a left adjoint~$F:\Set\to\ptSet$ which to every
set~$A$ associates the free pointed set it generates, that is the pointed set
$(A\uplus\{*\},*)$ where~$\uplus$ denotes the disjoint union (we often use the
notation~$*$ for the newly added element). We write~$\fpointed=G\circ F$ for the
monad on~$\Set$ induced by this adjunction. It is well-known~\cite{maclane:cwm}
that

\begin{proposition}
  \label{prop:pset}
  The category of sets and partial functions is isomorphic to the Kleisli
  category~$\Set_{\fpointed}$ associated to the monad~$\fpointed$
  on~$\Set$. Moreover, this Kleisli category is equivalent to the
  category~$\pointed\Set$.
\end{proposition}
\begin{proof}
  A partial function~$f:A\to B$ induces a morphism~$g:A\to B$
  in~$\Set_{\fpointed}$ (\ie a morphism $g:A\to\fpointed B$ in~$\Set$) defined
  on every $x\in A$ by $g(x)=f(x)$ if $f(x)$ is defined and $g(x)=\ast$
  otherwise, where $\fpointed B=B\uplus\{*\}$. Conversely, any morphism $g:A\to
  B$ in $\Set_{\fpointed}$ (\ie morphism $g:A\to\fpointed B$ in~$\Set$, with
  $\fpointed B=B\uplus\{*\}$) induces a partial function $f:A\to B$ defined on
  every $x\in A$ such that $g(x)\neq *$ by $f(x)=g(x)$. These two operations can
  easily be shown to be inverse of each other, thus exhibiting an isomorphism
  between the category of sets and partial functions and the Kleisli
  category~$\Set_{\fpointed}$.

  By general properties of monads (see \cite{maclane:cwm}, exercises p.~144),
  the category~$\Set_{\fpointed}$ is equivalent to the full subcategory
  of~$\pointed\Set$ whose objects are of the form~$FA$ for some set~$A\in\Set$.
  Moreover, every object $(A,a)$ of~$\pointed\Set$ is isomorphic to the pointed
  set~$F(A\setminus\{a\})$. The categories~$\Set_{\fpointed}$ and~$\pointed\Set$
  are thus equivalent.
\end{proof}

\noindent
The proposition above formalizes the fact that a partial function~$f:A\to B$ can
be seen as a total function~$f:A\to B\uplus\{*\}$ where~$f$ is ``undefined'' on
an element \hbox{$a\in A$} whenever~$f(a)=*$. The second part of the proposition
states that this partial function can also be seen as a pointed function
\hbox{$f:(A\uplus\{*\},*)\to(B\uplus\{*\},*)$}.

The situation between precubical sets and cubical sets is very similar. There is
an obvious inclusion functor $\pccat\to\ccat$, which by precomposition, induces
a forgetful functor~$U:\CSet\to\PCSet$ on the corresponding presheaf
categories. By general theorems (see Section~\ref{sec:cs-relate}), this functor
admits a left adjoint \hbox{$F:\PCSet\to\CSet$}. As previously, we write
$\fpointed=G\circ F$ for the induced monad on~$\PCSet$ and~$\pPCSet$ for the
Kleisli category associated to the monad. The morphisms in $\pPCSet$ should be
thought as ``partial morphisms of precubical sets''. And actually, this category
can be shown to be isomorphic to a category whose objects are precubical sets
and morphisms $\kappa:C\to D$ are families $(k_n:C(n)\to D(n))_{n\in\N}$ of
\emph{partial} functions satisfying suitable properties, which we do not need to
detail here.

One of the main interests of expressing the ``partial'' variants of models as
Kleisli constructions is that this enables us to easily lift the adjunctions
between models into adjunctions between their partial variants. Namely,

\begin{proposition}
  \label{prop:adj-lift}
  Suppose that~$\C$ and~$\D$ are categories and with~$S$ and~$T$ monads on
  respectively~$\C$ and~$\D$. Suppose moreover that $U:\D\to\C$ is a functor
  such that
  \[
  U\circ T=S\circ U
  \]
  and~$F$ sends the unit and the multiplication of~$T$ to the unit and the
  multiplication of~$U$. Then~$U$ has a left adjoint if and only if the
  functor~$U\circ I_D:\D_T\to\C_S$ has a left adjoint, where $I_D:\D_T\to\D$ is
  the canonical comparison functor between the Kleisli category~$\D_T$
  associated to~$T$ and~$\D$.
\end{proposition}

\noindent
This property, which is proved in a more general version
in~\cite{mulry:lifting-kleisli}, thus enables us to \emph{lift} an adjunction
between the categories~$\C$ and~$\D$ into an adjunction between the
corresponding Kleisli categories~$\C_S$ and~$\D_T$. In the following, it will be
particularly useful to lift adjunction between models into adjunctions between
corresponding models with partial morphisms.

\subsection{Symmetric cubical sets}
One sometimes needs more structure on cubical sets in order to formally express
the fact that the cells of dimension~$n\geq 2$ in cubical sets arising as models
for concurrent processes are essentially not directed. This can be formalized by
adjoining a notion of symmetry in cubical sets. The idea here is that given a
2-cell~$z$ in a cubical set as shown on the left of
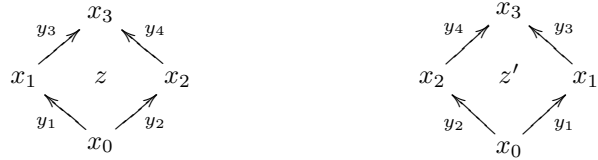
\[
\svxym{
  &x_3&\\
  x_1\ar[ur]^{y_3}&z&\ar[ul]_{y_4}x_2\\
  &\ar[ul]^{y_1}x_0\ar[ur]_{y_2}&\\
}
\qquad\qquad\qquad\qquad
\svxym{
  &x_3&\\
  x_2\ar[ur]^{y_4}&z'&\ar[ul]_{y_3}x_1\\
  &\ar[ul]^{y_2}x_0\ar[ur]_{y_1}&\\
}
\]
there should also be a ``mirror'' cell~$z'$ as shown on the right, expressing
the fact that~$z$ is not really directed from~$y_1y_3$ to~$y_2y_4$. The symmetry
of the cubical category will associate to each two cell a ``mirror'' 2-cell in
this way (it actually also generalizes this principle to higher dimensions). The
need for symmetry is also explained in the case of labeled cubical sets in
Example~\ref{ex:lcs}.

The \emph{symmetric cubical category} $\sccat{}$ is the free symmetric monoidal
category containing a co-cubical object. The presheaves on this category are
called \emph{symmetric cubical sets} and they form a category~$\SCSet$.
The category~$\sccat{}$ can also be described as the free monoidal category
containing a symmetric co-cubical
object~$(C,\varepsilon^-,\varepsilon^+,\eta,\gamma)$, which is a co-cubical
object $(C,\varepsilon^-,\varepsilon^+,\eta)$ together with a morphism
$\gamma:C\otimes C\to C\otimes C$ satisfying usual axioms for symmetry
\begin{equation}
  \label{eq:symmetry}
  (\gamma\otimes C)\circ(C\otimes\gamma)\circ(\gamma\otimes
  C)=(C\otimes\gamma)\circ(\gamma\otimes C)\circ(C\otimes\gamma)
  \qquad\qquad
  \gamma\circ\gamma=\gamma
\end{equation}
and
\[
\begin{array}{c@{\qquad}c@{\qquad}c}
  \gamma\circ(\varepsilon^-\otimes C)=C\otimes\varepsilon^-&\gamma\circ(\varepsilon^+\otimes C)=C\otimes\varepsilon^+&(\eta\otimes C)\circ\gamma=C\otimes\eta\\
  \gamma\circ(C\otimes\varepsilon^-)=\varepsilon^-\otimes C&\gamma\circ(C\otimes\varepsilon^+)=\varepsilon^-\otimes C&(C\otimes\eta)\circ\gamma=\eta\otimes C
\end{array}
\]
see~\cite{grandis-mauri:cubical-sets} for the details. Alternatively, the notion
of symmetric cubical set can be equivalently reformulated as a cubical set~$C$
together with, for every integer~$n$, an action of the symmetric
group~$\Sigma_n$ on~$C(n)$ -- the action of the transposition being given
by~$C(\gamma):C(2)\to C(2)$ -- which satisfies the following coherence axioms:
for every integers~$n$ and~$i$ such that~$0\leq i\leq n$ and
every~$\alpha\in\{-,+\}$,
\begin{itemize}
\item for every $(n+1)$\nbd{}cell~$x$ and permutation $\sigma\in\Sigma_{n+1}$,
  $\partial_i^\alpha(\sigma x)=\partial_{\sigma(i)}^\alpha(x)$
\item for every $n$\nbd{}cell~$x$ and permutation $\sigma\in\Sigma_n$,
  $\iota_i(\sigma x)=\iota_{\sigma(i)}(x)$
\end{itemize}
Namely, any symmetry \hbox{$\sigma:n\to n$} (\ie a bijection on a set with $n$
elements) can be decomposed as a product of transpositions and can therefore be
seen as a morphism in~$\sccat$ by sending the transposition $\sigma_i:n\to n$,
which exchanges the $i$-th and $(i+1)$\nbd{}th element, to the morphism
\hbox{$i\otimes\gamma\otimes(n-i-2)$}. The axioms~\eqref{eq:symmetry} imposed
on~$\gamma$, as well as the axioms of monoidal categories, ensure that this
operation is well defined. In the following, we will thus sometimes implicitly
consider a bijection as a morphism in the category~$\sccat$.
Given a symmetric monoidal category~$\C$ (such as~$\Set$ with cartesian
product), any cubical object of the underlying monoidal category of~$\C$ can be
canonically equipped with a structure of symmetric cubical set, the
morphism~$\gamma$ being given by the symmetry of the category.

Given an integer~$n$, we write~$\nsccat{n}$ for the full subcategory of~$\sccat$
whose objects are integers~$k\leq n$ and~$\nSCSet{n}$ for the category of
presheaves on~$\nsccat{n}$, whose objects are called \emph{$n$-dimensional
  symmetric cubical sets}. The \emph{symmetric precubical category} $\spccat$ is
defined similarly as the free symmetric monoidal category containing a
co-precubical object and we write~$\SPCSet$ for the category of presheaves on
$\spccat$, whose objects are called \emph{symmetric precubical sets}.  Notice
that many of the usual models for concurrency can be equipped with a similar,
and often related, notion of symmetry: for instance event
structures~\cite{winskel2007event,staton2010expressivity}, or Petri
nets~\cite{haymansymmetry}.

\subsection{Labeled cubical sets}
\label{sec:lcs}
We have explained that the 1-cells of a cubical set can be seen as occurrences
of events in the semantics of a concurrent computational process. One sometimes
needs to remember to which instruction of the process it corresponds. Labeled
(pre)cubical sets formally allows this. The presentation given here is adapted
from~\cite{goubault2001labelled}, see also~\cite{gaucher2010combinatorics}.

Suppose that we are given a set~$L$ of \emph{labels}. The
category~$(\Set,\times,1)$ has finite products and is thus monoidal with the
cartesian product as tensor and the terminal set~$1=\{*\}$ as unit (for
simplicity, we consider that the monoidal structure is strict). The set~$L$ can
be canonically equipped with a structure of symmetric precubical
object~$(L,\varepsilon^-,\varepsilon^+,\gamma)$ where
$\varepsilon^-,\varepsilon^+:L\to 1$ are both the terminal arrow and
$\gamma:L\times L\to L\times L$ is the canonical transposition. According to the
preceding remarks, it thus induces a symmetric precubical set noted~$\lcs L$ and
called the \emph{labeling precubical set on~$L$}. Moreover, if~$L'$ is another
set of labels, any function~$f:L\to L'$ induces a morphism between the
corresponding co-precubical objects, and therefore induces a morphism~$\lcs
f:\lcs L\to\lcs L'$, extending this operation into a functor. An explicit
description of the precubical set~$\lcs L$ can be given as follows: its
$n$\nbd{}cells~$l\in\lcs L(n)$ are lists \hbox{$l=(e_i)_{0\leq i<n}$}, of
length~$n$, of labels~$e_i\in L$. The face maps
$\partial_n^-,\partial_n^+:\lcs L(n+1)\to\lcs L(n)$ both send an
\hbox{$(n+1)$}\nbd{}cell $(e_i)_{0\leq i<n+1}$ to the list obtained by removing
the element at the $k$\nbd{}th position and the action of a
symmetry~$\sigma:n\to n$ on $\slcs L(n)$ sends a cell~$(e_i)_{0\leq i<n}$
to~$(e_{\sigma(i)})_{0\leq i<n}$.

It can be shown that~$\lcs L$ is the cofree precubical set generated by~$L$ in
the following sense:
\begin{proposition}
  \label{prop:lcs-cofree}
  The functor~$E:\SPCSet\to\Set$, which to every precubical set~$C$ associates
  its set~$C(1)/\!\approx$ of events (see Section~\ref{sec:cset}) and to every
  morphism~$\kappa:C\to D$ associates the
  function~$(\kappa_1/\approx):(C(1)/\!\approx)\to(D(1)/\!\approx)$, admits~$\lcs$ as
  right adjoint.
\end{proposition}
\begin{proof}
  Suppose given a precubical set~$C$ and a set~$L$. To every given function
  $f:(C(1)/\approx)\to L$, we associate the morphism~$\psi(f):C\to\lcs L$
  defined on an $n$\nbd{}cell~$x$ as
  the $n$-cell $(f(\partial_{\lnot
    0}^-(x)),\ldots,f(\partial_{\lnot{(n-1)}}^-(x)))$ of $\lcs L$, where the
  function~$\partial_{\lnot i}^\alpha$ is defined in Section~\ref{sec:mon-cc}.
  Conversely, to every morphism \hbox{$\kappa:C\to\lcs L$} of cubical sets, we
  associate the function $\psi(\kappa):(C(1)/\approx)\to L$ defined
  as~$\kappa_1/\!\approx$: this is well defined since the events of~$\lcs L$ are
  (in bijection with) the elements of~$L$.
  Finally, is it straightforward to check that the functions~$\varphi$
  and~$\psi$ are natural in~$C$ and~$L$, and inverse of each other.
\end{proof}

\begin{remark}
  The cofree non-symmetric labeling precubical set on a set~$L$ could be defined
  in the same way, but a direct description is more difficult. It can for
  example be obtained from the symmetric labeling precubical set~$\lcs L$ on~$L$
  by quotienting by the action of symmetries.
\end{remark}

\newcommand{\E}{\mathcal{E}}

Recall that given categories~$\C$, $\D$ and~$\E$ and functors~$F:\C\to\E$ and
$G:\D\to\E$, the \emph{slice category} $\slicecat{F}{G}$ (sometimes also called
\emph{comma category}) is the category whose objects are triples $(A,f,A')$
where~$A$ is an object of~$\C$, $A'$ is an object of~$\D$ and~$f:FA\to GA'$ is a
morphism of~$\E$, and whose morphisms $(h,h'):(A,f,A')\to(B,g,B')$ are the pairs
of morphisms \hbox{$h:A\to B$} of~$\C$ and $h':A'\to B'$ of~$\D$ making the
diagram
\[
\vxym{
  FA\ar[d]_f\ar[r]^{Fh}&FB\ar[d]^g\\
  GA'\ar[r]_{Gh'}&GB'\\
}
\]
commute. By abuse of notation, we often write~$\slicecat\D{G}$ for the
category~$\slicecat{\Id_\D}{G}$. A labeled variant of cubical sets is defined as
follows.

\begin{definition}
  The category of \emph{labeled symmetric precubical sets}, denoted by
  $\LSPCSet$, is the slice category~$\slicecat\SPCSet\lcs$.
\end{definition}

\noindent
By Proposition~\ref{prop:lcs-cofree}, a given labeled symmetric precubical set
$(C,\ell,L)$ (defined by \hbox{$C\in\SPCSet$}, $L\in\Set$ and $\ell:C\to\lcs L$) can
also be seen as a triple $(C,\ell,L)$ with the function $\ell:E(C)\to L$
associating a label to each event of~$C$. In other words, the
category~$\LSPCSet$ is isomorphic to~$\slicecat{E}{\lcs}$, where~$E$ is the
event functor introduced in Proposition~\ref{prop:lcs-cofree}.

\begin{example}
  \label{ex:lcs}
  The CCS processes~$ab+ba$ and~$(a|b)$ respectively induce labeled symmetric
  cubical sets of the form
  \[
  \svxym{
    &z&\\
    \ar[ur]^by_1&&y_2\ar[ul]_a\\
    &\ar[ul]^ax\ar[ur]_b&\\
  }
  \qquad\text{and}\qquad
  \svxym{
    &z&\\
    \ar[ur]^by_1\ar@{}[rr]|-{\alpha\ \beta}&&y_2\ar[ul]_a\\
    &\ar[ul]^ax\ar[ur]_b&\\
  }
  \]
  with in the second case two squares~$\alpha$ and~$\beta$ attached in the
  middle, respectively labeled by~$(a,b)$ and~$(b,a)$ (and none in the first
  case): the presence of a square indicates that the two actions~$a$ and~$b$
  commute and more generally $n$-cubes indicate the commutation of~$n$
  actions~\cite{EGGeometry}. Notice that the symmetry intuitively enables us to
  say that the cell labeled by~$ab$ is ``the same as'' the cell labeled
  by~$ba$. In a non-symmetric case, there would be only one cell and there is no
  canonical choice of naming for this cell (this is however sometimes overcome
  by supposing that letters are totally ordered, but supposing this is not very
  natural).
\end{example}

\noindent
Notice that events provides a canonical labeling of symmetric precubical sets:

\begin{proposition}
  The forgetful functor $U:\LSPCSet\to\SPCSet$ which to every labeled symmetric
  precubical set associates the underlying symmetric precubical set (forgetting
  the labels) admits the functor \hbox{$E:\SPCSet\to\LSPCSet$} as left adjoint,
  which to every symmetric precubical set~$C$ associates the labeled symmetric
  precubical set~$(C,\ell,\lcs L)$ where~$L=C(1)/\!\approx$ is the set of events
  of~$C$ and~$\ell$ is the morphism induced by the function~$\ell:C(1)\to L$
  which to every 1-cell associates its equivalence class under~$\approx$.
\end{proposition}

\noindent
This means in particular that all following results about labeled symmetric
precubical sets simply extend to the unlabeled case by considering precubical
sets labeled by their events. We thus only handle labeled cases in the
following, since unlabeled structures are a particular instance.

The category of labeled symmetric cubical sets is defined in a similar way. A
given pointed set~$(L,*)$ induces a symmetric cubical object
$(L,\varepsilon^-,\varepsilon^+,\eta,\gamma)$ where
$\varepsilon^-,\varepsilon^+:L\to 1$ are both the terminal arrow, $\eta:1\to L$
associates~$*$ to the unique element of~$1$ and $\gamma:L\times L\to L\times L$
is the canonical transposition function. As previously, this induces a symmetric
cubical set, that we still write~$\lcs(L,*)$, and can be shown to be cofree in
the sense that

\begin{proposition}
  The functor~$E:\SCSet\to\pointed\Set$, which to every cubical set~$C$
  associates the pointed set obtained from~$C(1)/\!\approx$ by identifying all
  equivalence classes containing an element of the image of~$\iota_0$ to a
  single element~$*$ chosen as distinguished element and to every
  morphism~$\kappa:C\to D$ associates the morphism induced by $\kappa_1:C(1)\to
  D(1)$, admits~$\lcs$ as right adjoint.
\end{proposition}

\begin{definition}
  The category of \emph{symmetric labeled cubical sets}, denoted by $\LSCSet$,
  is the slice category~$\slicecat\SCSet\lcs$, which is isomorphic to
  $\slicecat{E}{\lcs}$.
\end{definition}

\noindent
An explicit description of~$\lcs(L,*)$ is similar to the one of the labeling
symmetric precubical set: its $n$-cells are lists $l=(e_i)_{0\leq i<n}$, with
the same face and symmetry maps as previously. The degeneracy maps
$\iota_i:\lcs(L,*)(n)\to\lcs(L,*)(n+1)$ associate to every list~$l$ of
length~$n$ the list of length $n+1$ obtained from~$l$ by inserting~$*$ at the
$i$-th position.

We have defined labellings in the most natural way. There is however a slight
mismatch between labeled precubical and cubical sets: in the first case
functions between labels are total whereas they are partial in the second
case. This mismatch actually turns out to bring annoying details, as explained
in Section~\ref{sec:cs-relate} (see also~\cite{fahrenberg2005category}). The
opposite choices can be made in both cases as follows. A slightly more general
notion of labeled precubical set can be defined, by allowing partial functions
between morphisms. If we write $U:\pointed\Set\to\Set$ for the canonical
forgetful functor, the category of \emph{weakly labeled symmetric precubical
  sets}~$\wLSPCSet$ is defined as~$\wLSPCSet=\slicecat{\SPCSet}{\lcs
  U}$. Conversely, one can restrict labeled cubical sets by only allowing total
functions between labels and imposing that only degenerate events are labeled by
the distinguished element of the labeling pointed set thus defining a category
of \emph{totally labeled symmetric cubical sets} (we do not detail this
construction here).

Finally, we introduce the notion of strongly labeled cubical set, which will
turn out in Section~\ref{sec:adjunctions} to be the ``right'' notion of labeled
cubical set in order to relate them with most of the usual models of
concurrency.

\begin{definition}
  A labeled cubical set~$(C,\ell)$ is \emph{strongly labeled} when there exists
  no pair of distinct $k$-cells, for some dimension $k$, whose sources and
  targets are equal, which have the same label: for every index~$k>0$, and every
  elements $x,y\in C(k)$ such that for every index~$0\leq i<k$
  $\partial_i(x)=\partial_i(y)$, if $\ell(x)=\ell(y)$ then $x=y$.
\end{definition}

\noindent
This condition can be seen as a labeled and higher dimensional analogue of
Winskel's ``no ravioli'' condition for HDA~\cite{winskel2007event}, which
imposes that two parallel $1$-cells should be equal, and corresponds to being
separated \wrt a Grothendieck topology.

\subsection{Higher dimensional automata}
A \emph{pointed cubical set} $(C,i)$ is a cubical set together with a
distinguished $0$\nbd{}cell \hbox{$i\in C(0)$}. The notion of higher dimensional
automaton can be seen as a generalization of the classical notion of automaton
to higher dimensional transition systems:

\begin{definition}
  A \emph{higher dimensional automaton} (or \emph{HDA}) is a pointed labeled
  symmetric cubical set~$C$, the distinguished element~$i$ being called the
  \emph{initial state}. A morphism of HDA is a morphism between the underlying
  labeled symmetric cubical sets which preserves the initial state.
\end{definition}

\noindent
Given a category~$\C$ of cubical sets, we often write~$\pointed\C$ for the
corresponding category of pointed cubical sets. We write~$\HDA=\pointed\LSCSet$
for the category of HDA and \hbox{$\sHDA=\pointed\LSPCSet$} for the category of
\emph{strict HDA}. We also write~$\nHDA{n}=\pointed{\nLSCSet{n}}$ for the
subcategories for truncated HDA.

A \emph{path} $p:x\transitionpath{}x'$ in an HDA~$C$ is a finite
sequence~$(y_i)_{0\leq i<n}$ of 1-cells of~$C$ such
that~$\partial_0^+(y_i)=\partial_0^-(y_{i+1})$, $\partial_0^-(y_0)=x$ and
$\partial_0^+(y_n)=x'$. We write $s\cdot t$ for the concatenation of two
paths~$s$ and~$t$. A 0-cell~$x$ of an HDA is \emph{reachable} when there exists
a path~$s:i\transitionpath{}x$, where~$i$ is the initial state of the HDA. Since
higher dimensional cells express the fact that transitions are independent, two
paths differing only by a reordering of independent transitions should be
considered as equivalent from the concurrency point of view. This is formally
expressed by the \emph{homotopy} relation between
paths~\cite{fajstrup2005dipaths,van2006expressiveness}, which is defined as the
smallest equivalence relation relating two paths $s\cdot m\cdot n\cdot t$ and
$s\cdot p\cdot q\cdot t$ where~$m$, $n$, $p$ and $q$ are 1-cells such that there
exists a 2-cell $z$ for which $m=\partial_0^-(z)$, $q=\partial_0^+(z)$,
$p=\partial_1^-(z)$ and $n=\partial_1^+(z)$ ; graphically,
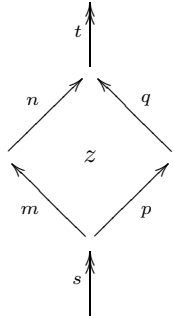
\[
\vxym{
  &&\\
  &\ar@{->>}[u]^t&\\
  \ar[ur]^n&z&\ar[ul]_q\\
  &\ar[ul]^m\ar[ur]_p\\
  &\ar@{->>}[u]^s\\
}
\]
In particular, in the situation above, $m$ and $q$ (\resp $p$ and $n$) are part
of the same event. Given two paths~$s$ and~$t$, we write $s\sim t$ when they are
homotopic. Two homotopic paths are necessarily parallel (they have the same
source and target).

\subsection{Relating variants of cubical sets.}
\label{sec:cs-relate}
Suppose given two categories~$\C$ and~$\D$ and a functor
\hbox{$I:\C\to\D$}. Every presheaf \hbox{$C:\D^\op\to\Set$} on~$\C$ induces a
presheaf \hbox{$C\circ I^\op:\C^\op\to\Set$} by precomposition with~$I$, and
this operation extends into a functor \hbox{$\hat{I}:\hat\D\to\hat\C$} from the
presheafs on~$\D$ to those on~$\C$, defined on morphisms~$\alpha:C\to D$ by
$(\hat I(\alpha))_A=\alpha_{I(A)}$. These functors have many nice properties,
some of which useful here are detailed below:

\begin{proposition}
  \label{prop:psh-adj}
  Suppose given two categories~$\C$ and~$\D$, where~$\C$ is small, and a functor
  $I:\C\to\D$ between them.
  \begin{enumerate}
  \item The functor~$\hat{I}:\hat\D\to\hat\C$ admits both a left and a right
    adjoint and we write~$T$ for the monad induced on~$\hat\C$.
  \item The Kleisli category~$\D_T$ associated to the monad~$T$ embeds fully and
    faithfully into~$\hat\D$.
  \item When~$I$ is bijective on objects, the adjunction is monadic which means
    that the category~$\hat\D$ is equivalent to the category~$\C^T$ of algebras
    for the monad~$T$ on~$\hat\C$.
  \end{enumerate}
\end{proposition}
\begin{proof}
  (i) and (ii) are standards properties~\cite{mac1992sheaves}. In particular,
  the free presheaf in~$\hat\D$ on a presheaf~$C\in\hat\C$ can be computed as
  the left Kan extension of~$C$ along~$I$ (and similarly for the right adjoint).

  (iii) This fact does not seem to be very well-known and can be found for
  example p.~105 of~\cite{borceux2001galois}. We have seen that the
  functor~$\hat I$ admits a left adjoint. Since it is the bijective on objects
  it is conservative (it reflects isomorphisms): an isomorphism between
  presheaves is simply a natural transformation between them whose components
  are all invertible. Moreover, presheaf categories are cocomplete; in
  particular, they have all equalizers, these are computed pointwise and they
  are thus preserved by precomposition with~$U$. We can conclude by using Beck's
  monadicity theorem~\cite{maclane:cwm}.
\end{proof}

\noindent
Notice that this generalizes in particular the situation described in
Section~\ref{sec:partial-pcset}.

This property is very interesting because, it means that all the forgetful
functors between variants of categories of cubical sets admit both left and
right adjoint:
\begin{itemize}
\item functors forgetting structure:
  \[
  \begin{array}{c}
    \SCSet\to\CSet
    \qcomma
    \CSet\to\PCSet
    \qcomma
    \PCSet\to\Set
    \qcomma
    \text{etc.}
    \\
    \SCSet_n\to\CSet_n
    \qcomma
    \CSet_n\to\PCSet_n
    \qcomma
    \PCSet_n\to\Set
    \qcomma
    \text{etc.}
  \end{array}
  \]
\item truncation functors:
  \[
  \SCSet\to\SCSet_n
  \qcomma
  \CSet\to\CSet_n
  \qcomma
  \text{etc.}
  \]
\end{itemize}

\noindent
These adjoints will allow us to compute for example the free cubical set on a
precubical set and so on, and will be used in the following.
As an illustration, consider the functor~$\PCSet\to\nPCSet{n}$. Given an
$n$\nbd{}dimensional precubical set~$C$, the left adjoint sends~$C$ to the
precubical set~$D$ whose $k$-cells are \hbox{$D(k)=C(k)$} for $k\leq n$
and~$D(k)=\emptyset$ otherwise. The action of the right adjoint is more subtle:
it sends~$C$ to the precubical set obtained from~$C$ by ``filling in'' all the
$k$-dimensional cubes, with~$k>n$, by a $k$-cell.


The analogy between the adjunction between sets and pointed sets and the
adjunction between precubical sets and sets, can be related with the
construction of labeling cubical sets as follows.

\begin{lemma}
  The diagram
  \[
  \vxym{
    \Set\ar@/^/[r]^F\ar[d]_I\ar@{}[r]|-{\top}&\ar@/^/[l]^E\SPCSet\ar[d]^J\\
    \pointed\Set\ar@/^/[r]^H\ar@{}[r]|-{\bot}&\ar@/^/[l]^G\SCSet
  }
  \]
  commutes, in the sense that $J\circ F=H\circ I$ and~$G\circ J=I\circ E$,
  where~$E$ (\resp $H$) is the functor which to every symmetric precubical
  (\resp cubical set) associates its set (\resp pointed set) of events described
  along with their right adjoints in Section~\ref{sec:lcs}, and $I$ (\resp~$J$) is
  the left adjoint to the forgetful functor~$\pointed\Set\to\Set$
  (\resp~$\SCSet\to\PCSet$).
\end{lemma}

\noindent
It could be hoped that previous Lemma would provide the starting point of a
lifting of the adjunctions between~$\SPCSet$ and~$\SCSet$ to adjunctions
between~$\LSPCSet$ and~$\LSCSet$. However this is not the case: morphisms
between labels have to be total or partial in both the categories. It is
however easy to show that

\begin{proposition}
  The forgetful functor $\LSCSet\to\wLSPCSet$ admits both a left and a right
  adjoint (and other adjunctions mentioned above can be lifted to the labeled
  case in a similar way). Similar adjunctions also exist between the variants
  where functions between labels are total.
\end{proposition}

\noindent
The choice of partial or total functions between labels in the category of
labeled symmetric (pre)cubical sets is thus difficult to handle in a modular
way. The choice has to be made once for all and in the following, we
deliberately do not explicit which one is made since all the constructions given
here work in both cases.

\section{Traditional models for concurrency}
\label{sec:traditional-models}
\subsection{Transition systems}
\label{TS}
Transition systems are one of the oldest semantic models, both for sequential
and concurrent systems, in which computations are modeled as the sequence of
interactions that they can have with their environment. There is a convenient
categorical treatment of this model, that we use in the sequel, taken from
\cite{winskel-nielsen:models-concur}.

\begin{definition}
  A \emph{transition system} is a quadruple $(S,i,E,Tran)$ where
  \begin{itemize}
  \item $S$ is a set of \emph{states} with \emph{initial state} $i$,
  \item $E$ is a set of \emph{events},
  \item $Tran \subseteq S \times E\times S$ is the \emph{transition} relation.
  \end{itemize}
\end{definition}

\noindent
In other words, a transition system is a graph together with a distinguished
vertex.  Transition systems are made into a category by defining morphisms to be
some kind of simulation (for then being able to discuss about properties modulo
weak or strong bisimulation, see \cite{AJMNGWBisimulation}). The idea is that a
transition system $T_1$ simulates a transition system $T_0$ if as soon as $T_0$
can fire some action $a$ in some context, $T_1$ can fire $a$ as well in some
related context. A morphism $f: T_0 \to T_1$ defines the way states and
transitions of $T_0$ are related to states and transitions of $T_1$ making
transition systems into a category $\TS$.

\begin{definition}
  Let $T_0=(S_0,i_0,E_0,Tran_0)$ and $T_1=(S_1,i_1,E_1,Tran_1)$ be two
  transition systems. A \emph{partial morphism}
  $f:T_0 \rightarrow T_1$ is a pair $f=(\sigma,\tau)$ where $\sigma: S_0
  \rightarrow S_1$ is a function and $\tau: E_0 \rightarrow E_1$ is a
  \emph{partial} function such that
  \begin{itemize}
  \item $\sigma(i_0)=i_1$,
  \item $ (s,e,s') \in Tran_0$ and $\tau(e)$ is defined implies
    $(\sigma(s),\tau(e),\sigma(s')) \in Tran_1 $. Otherwise, if $\tau(e)$
    is undefined then $\sigma(s)=\sigma(s')$.
  \end{itemize}
\end{definition}

As in~\cite{winskel-nielsen:models-concur}, we can restrict to \emph{strict}
morphisms, i.e.~the ones for which $\tau$ is a total function, by suitably
completing transition systems.  Partial morphisms can then be recovered by
adding ``idle'' transitions to the systems, similarly to the construction of the
category of sets and partial functions as the Kleisli category associated to the
free pointed set monad~$\fpointed$ on~$\Set$ given in
Section~\ref{sec:partial-pcset}.

An idle transition is a transition~$*$ which goes from a state $s$ to the same
state~$s$. Consider the following completion $T_*=(S_*,i_*,E_*,Tran_*)$ of a
transition system $T=(S,i,E,Tran)$, by setting $S_*=S$, $i_*=i$,
$E_*=E\uplus\{*\}$ and \hbox{$Tran_*=Tran\uplus\{(s,*,s)\tq s \in S \}$}. Now,
by the preceding remarks a total morphism $(\sigma,\tau)$ from $(T_0)_*$ to
$(T_1)_*$ such that $\tau(*)=*$ is the same as a partial morphism from $T_0$
to~$T_1$.
Again, the operation $(-)_*$ induces a monad on the category~$\sTS$ of
transition systems and strict morphisms, and the category~$\TS$ can be recovered
as the Kleisli category associated to this monad. Likewise, all the models for
concurrency considered in this article admit a ``strict'' variant, from which
the ``non-strict'' model can be reconstructed by a Kleisli construction. For
lack of space we will not detail all the variants here.


\begin{example}
  \label{ex:lts}
  The CCS processes~$a\cdot(b+c)$, $a\cdot(b|c)$ and~$a\cdot(b\cdot c+c\cdot b)$
  respectively induce the following transition systems:
  \[
  \svxym{
    &y&\\
    \\
    &\ar@/^/[uu]^by\ar@/_/[uu]_c&\\
    &\ar[u]^ax&\\
  }
  \qquad\qquad\qquad
  \svxym{
    &z&\\
    y_1\ar[ur]^c&&y_2\ar[ul]_b\\
    &\ar[ul]^by\ar[ur]_c&\\
    &\ar[u]^ax&\\
  }
  \qquad\qquad\qquad
  \svxym{
    &z&\\
    y_1\ar[ur]^c&&y_2\ar[ul]_b\\
    &\ar[ul]^by\ar[ur]_c&\\
    &\ar[u]^ax&\\
  }
  \]
\end{example}

\subsection{Asynchronous automata}
Asynchronous automata are a nice generalization of both transition systems and
Mazurkiewicz traces, and have influenced a lot of other models for concurrency,
such as transition systems with independence (or asynchronous transition
systems). They have been independently introduced in \cite{MWSConcurrent} and
\cite{MABCategories}. The idea is to decorate transition systems with an
\emph{independence} relation between actions that will allow us to distinguish
between true-concurrency and mutual exclusion (or non-determinism) of two
actions. For example, the two last transition systems of Example~\ref{ex:lts} do
not allow us to distinguish between processes which are arguably different from
the concurrency point of view. We actually use a slight modification for our
purposes, due to \cite{MDRMSPetri}, called \emph{automaton with concurrency
  relations}:

\begin{definition}
  \label{def:acr}
  An \emph{automaton with concurrency relations}~\hbox{$(S,i,E,Tran,I)$} is a
  quintuple where
  \begin{itemize}
  \item $(S,i,E,Tran)$ is a transition system,
  \item $Tran$ is such that whenever $(s,a,s')$, $(s,a,s'') \in Tran$, then
    $s=s''$,
  \item $I=(I_s)_{s\in S}$ is a family of irreflexive, symmetric binary
    relations $I_s$ on~$E$ such that whenever we have $a_1\indep{s}a_2$ (with
    $a_1,a_2 \in E$), there exist transitions $(s,a_1,s_1)$, $(s,a_2,s_2)$,
    $(s_1,a_2,r)$ and $(s_2,a_1,r)$ in $Tran$.
  \end{itemize}
\end{definition}

\noindent A morphism of automata with concurrency relations consists of a
morphism \hbox{$(\sigma,\tau)$} between the underlying transition systems such
that $ a\indep{s}b$ implies that $\tau(a)\indepp{\sigma(s)}\tau(b)$ whenever
$\tau(a)$ and $\tau(b)$ are both defined. This makes automata with concurrency
relations into a category, written $\ACR$. We also write~$\sACR$ for the variant
of this category where morphisms are strict morphisms. Again, the
category~$\sACR$ can be constructed from~$\ACR$ by a Kleisli construction, using
$*$-transitions and total morphisms (the condition on the independence relation
is then that $a \indep{s} b$ implies $\tau(a) \indepp{\sigma(s)} \tau(b)$
whenever $\tau(a)\neq *$ and $\tau(b)\neq *$).


\begin{example}
  The CCS processes $a\cdot(b|c)$ and~$a\cdot(b\cdot c+c\cdot b)$ induces the
  labeled asynchronous transition systems whose underlying transition system are
  isomorphic and shown in Example~\ref{ex:lts}. The independence relation
  contains~$e_bI_ye_c$ for the first process (where~$e_b$ and~$e_c$ are the
  events with source~$y$, labeled respectively by~$b$ and~$c$) and is empty for
  the second process.
\end{example}

\subsection{Event structures}
Event structures were introduced in~\cite{event1,event2} in order to abstract
away from the precise places and times at which events occur in distributed
systems. The idea is to focus on the notion of event and the causal ordering
between them. We recall below the definition of (unlabeled prime) event
structures.
\begin{definition}
  An \emph{event structure}~$(E,\leq,\incompat)$ consists of a poset~$(E,\leq)$
  of \emph{events}, the partial order relation expressing \emph{causal
    dependency}, together with a symmetric irreflexive relation $\incompat$
  called \emph{incompatibility}
  satisfying
  \begin{itemize}
  \item finite causes: for every event $e$, the set $\setof{e'\tq e' \leq e}$ is
    finite,
  \item hereditary incompatibility: for every events $e$, $e'$ and $e''$, $e
    \incompat e'$ and $e' \leq e''$ implies $e \incompat e''$.
  \end{itemize}
\end{definition}

\noindent
We write~$\ES$ for the category of event structures, a morphism between two
event structures $(E,\leq,\incompat)$ and~$(E',\leq',\incompat')$ consisting of
a partial function $f:E\to E'$ which is such that
\begin{itemize}
\item if $f(e)$ is defined then $\setof{e' \tq e' \leq f(e)} \subseteq
  f(\setof{e'' \tq e'' \leq e})$,
\item and if $f(e_0)$ and $f(e_1)$ are both defined and we have either $f(e_0)
  \incompat' f(e_1)$ or $f(e_0)=f(e_1)$ then either $e_0 \incompat e_1$ or
  $e_0=e_1$.
\end{itemize}

A \emph{labeled event structure} consists of an event structure
$(E,\leq,\incompat)$ together with a set~$L$ of \emph{labels} and a
\emph{labeling function}~$\ell:E\to L$ which to every event associates a
label. A
morphism~$(f,\lambda):(E,\leq,\incompat,\ell,L)\to(E',\leq',\incompat',\ell',L')$
of labeled event structure consists of a
morphism~$f:(E,\leq,\incompat)\to(E',\leq',\incompat')$ between the underlying
event structures and a partial function~$\lambda:L\to L'$ between the sets of
labels such that~$\ell'\circ f=\lambda\circ\ell$. We write~$\LES$ for the
category of labeled event structures.
We also write~$\sES$ (\resp $\sLES$) for the category of \emph{strict} (labeled)
event structures, defined as the subcategory of~$\ES$ (\resp $\LES$) whose
morphisms are total functions -- these categories can also be obtained by
suitable Kleisli constructions.

\begin{example}
  The CCS processes~$a\cdot(b+c)$, $a\cdot(b|c)$ and~$a\cdot(b\cdot c+c\cdot b)$
  respectively induce the following labeled event structures (to be read from
  bottom up, the continuous lines representing the partial order and the dotted
  ones expressing incompatibilities):
  \[
  \svxym{
    \\
    b\ar@{.}[rr]&&c\\
    &\ar@{-}[ul]a\ar@{-}[ur]&
  }
  \qquad\qquad\qquad
  \svxym{
    \\
    b&&c\\
    &\ar@{-}[ul]a\ar@{-}[ur]&
  }
  \qquad\qquad\qquad
  \svxym{
    c\ar@{.}[rr]\ar@{.}[drr]&&\ar@{.}[dll]b\\
    b\ar@{-}[u]\ar@{.}[rr]&&c\ar@{-}[u]\\
    &\ar@{-}[ul]a\ar@{-}[ur]&
  }
  \]
  Notice that in the last one, $b$ and~$c$ appear twice: this is because we have
  figured the labels and not the events (and two distinct events can of course
  have the same label).
\end{example}

\subsection{Petri nets}
Petri nets are a well-known model of parallel computation, generalizing
transition systems by using a built-in notion of resource. This allows for
deriving a notion of independence of events, which is much more general than the
independence relation of asynchronous transition systems. They are numerous
variants of Petri nets
since they were introduced in~\cite{carl}, and we choose the definition used by
Winskel and Nielsen in~\cite{winskel-nielsen:models-concur}, since this is
well-suited for formal comparisons with other models for concurrency:

\begin{definition}
  A \emph{Petri net}~$N$ is a tuple $(P,M_0,E,\Pre,\Post)$ where
  \begin{itemize}
  \item $P$ is a set of \emph{places},
  \item $M_0\in\N^P$ is the \emph{initial marking},
  \item $E$ is a set of \emph{events},
  \item $\Pre:E\to\N^P$ and $\Post:E\to\N^P$ are the \emph{precondition} and
    \emph{postcondition} functions.
\end{itemize}
\end{definition}
\noindent
When there is no ambiguity, given an event~$e$ of a Petri net~$N$, we often
write~$\pre{e}$ for~$\Pre(e)$ and~$\post{e}$ for~$\Post(e)$. A
\emph{marking}~$M$ is a function in~$\N^P$, which associates to every place the
number of resources (or tokens) that it contains.
The sum $M_1+M_2$ of two markings~$M_1$ and~$M_2$ is their pointwise sum.
An event~$e$ induces a \emph{transition} between two markings~$M_1$ and~$M_2$,
that we write $M_1\transition{e}M_2$, whenever there exists a marking~$M$ such
that $M_1=M+\pre{e}$ and $M_2=M+\post{e}$.

A \emph{morphism of Petri nets} $(\varphi,\psi):N\to N'$, between the two Petri
nets \hbox{$N=(P,M_0,E,\Pre,\Post)$} and \hbox{$N'=(P',M_0',E',\Pre',\Post')$},
consists of a function \hbox{$\varphi:P'\to P$} and a partial
function~$\psi:E\to E'$ such that for every place $p\in P'$ and event
\hbox{$e\in E$},
$M_0'=M_0\circ\varphi$, $\pre{\psi(e)}=\pre{e}\circ\varphi$ and
$\post{\psi(e)}=\post{e}\circ\varphi$.
We write~$\PNet$ for the category of Petri nets and~$\sPNet$ for the subcategory
whose morphisms have total functions on events.  Notice that the partial
function \hbox{$\varphi:P'\to P$} on places goes ``backwards''. This might seem
a bit awkward at first sight and we explain why this is the ``right'' notion of
morphism in Remark~\ref{rem:pn-morph-direction}.





A \emph{labeled Petri net} is a Petri net together with a set~$L$ of labels and
a function \hbox{$\ell:E\to L$} labeling events. The notion of morphism of Petri
nets can be extended in a straightforward way to labeled ones and we
write~$\LPNet$ for the category of labeled Petri nets and~$\sLPNet$ for the
subcategory whose morphisms are total functions.

\begin{example}
  The CCS processes~$a\cdot(b+c)$, $a\cdot(b|c)$ and~$a\cdot(b\cdot c+c\cdot b)$
  respectively induce the following labeled Petri nets:
    \[
  \scalebox{0.65}
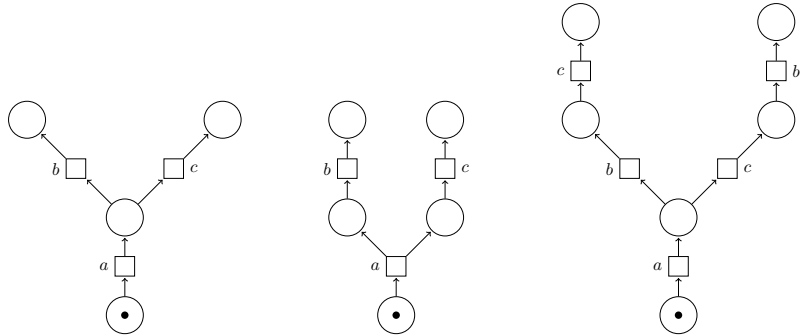
{
    \begin{tikzpicture}
      \node[place,tokens=1] (x) at (2,0) {};
      \node[place,tokens=0] (y) at (2,2) {};
      \node[place,tokens=0] (z1) at (0,4) {};
      \node[place,tokens=0] (z2) at (4,4) {};
      \node[transition,label=left:$a$] at (2,1) {}
      edge[pre] (x)
      edge[post] (y);
      \node[transition,label=left:$b$] at (1,3) {}
      edge[pre] (y)
      edge[post] (z1);
      \node[transition,label=right:$c$] at (3,3) {}
      edge[pre] (y)
      edge[post] (z2);
    \end{tikzpicture}
    \qquad\qquad
    \begin{tikzpicture}
      \node[place,tokens=1] (x) at (2,0) {};
      \node[place,tokens=0] (y1) at (1,2) {};
      \node[place,tokens=0] (y2) at (3,2) {};
      \node[place,tokens=0] (z1) at (1,4) {};
      \node[place,tokens=0] (z2) at (3,4) {};
      \node[transition,label=left:$a$] at (2,1) {}
      edge[pre] (x)
      edge[post] (y1)
      edge[post] (y2);
      \node[transition,label=left:$b$] at (1,3) {}
      edge[pre] (y1)
      edge[post] (z1);
      \node[transition,label=right:$c$] at (3,3) {}
      edge[pre] (y2)
      edge[post] (z2);
    \end{tikzpicture}
    \qquad\qquad
    \begin{tikzpicture}
      \node[place,tokens=1] (x) at (2,0) {};
      \node[place,tokens=0] (y) at (2,2) {};
      \node[place,tokens=0] (z1) at (0,4) {};
      \node[place,tokens=0] (z2) at (4,4) {};
      \node[place,tokens=0] (t1) at (0,6) {};
      \node[place,tokens=0] (t2) at (4,6) {};
      \node[transition,label=left:$a$] at (2,1) {}
      edge[pre] (x)
      edge[post] (y);
      \node[transition,label=left:$b$] at (1,3) {}
      edge[pre] (y)
      edge[post] (z1);
      \node[transition,label=left:$c$] at (0,5) {}
      edge[pre] (z1)
      edge[post] (t1);
      \node[transition,label=right:$c$] at (3,3) {}
      edge[pre] (y)
      edge[post] (z2);
      \node[transition,label=right:$b$] at (4,5) {}
      edge[pre] (z2)
      edge[post] (t2);
    \end{tikzpicture}
  }
  \]
  In the diagrams above, we have used the usual notation for Petri nets: square
  nodes represent transitions, circled ones represent places (with dots
  indicating tokens) and arrows represent pre- and postconditions.
\end{example}

\section{Relating models for concurrency}
\label{sec:adjunctions}

The purpose of this section is to relate traditional models introduced in
\Section{sec:traditional-models} with the geometric models of
\Section{sec:geometric-models} (mainly HDA).

\subsection{Transition systems and HDA}
In this section, we relate labeled transition systems and HDA. We begin by
relating transition systems to the category of $1$-dimensional HDA by defining
two adjoint functors
\[
F:\nsHDA{1}\to\sTS
\qqtand
G:\sTS\to\nsHDA{1}
\]

We define the functor~$F$ as follows. To a $1$-dimensional HDA~$C$ labeled
by~$L$, we associate the transition system $(S,i,E,Tran)$ defined by
\hbox{$S=C(0)$}, $i$~being the distinguished element of~$C$, $E=L$ and the
transitions being defined by
\hbox{$Tran=\setof{(\partial_0^-(e),\ell(e),\partial_0^+(e))\tq e\in E}$}. And
to any morphism \hbox{$(\varphi,\lambda):C\to D$} between labeled precubical
sets, we associate the morphism $(\sigma,\tau)$ which is defined by
\hbox{$\sigma=\varphi_0:C(0)\to D(0)$} and~$\tau=\lambda$. The functor is
defined in the obvious way on morphisms.

Conversely, the functor~$G$ is defined as follows. To any transition system
\hbox{$T=(S,i,E,Tran)$}, we associate the strict $1$-dimensional HDA~$C$ labeled
by~$E$ whose underlying precubical set~$C$ is such that $C(0)=S$,
\hbox{$C(1)=Tran$}, the face morphisms \hbox{$\partial_0^-:C_1\to C_0$} and
$\partial_0^+:C_1\to C_0$ are respectively defined by
\hbox{$\partial_0^-(s,e,s')=s$} and $\partial_0^+(s,e,s')=s'$, the labeling
function is defined by~$\ell(s,e,s')=e$ and the distinguished element is the
distinguished element \hbox{$i\in C(0)$}. To any morphism
\hbox{$(\sigma,\tau):(S_1,i_1,E_1,Tran_1)\to(S_2,i_2,E_2,Tran_2)$} we associate
the morphism~$(\kappa,\lambda)$ of HDA, where~$\kappa$ is the morphism of
pointed 1-dimensional precubical set whose components are~$\kappa_0=\sigma$
and~$\kappa_1=\tau$, the morphism~$\lambda$ between labels being~$\tau$. The
functor is defined in the obvious way on morphisms.

The functors defined above enable us to relate both models:

\begin{theorem}
  \label{thm:sts-adj}
  The functor $F:\nsHDA{1}\to\sTS$ defined above is left adjoint to the functor
  \hbox{$G:\sTS\to\nsHDA{1}$}. The comonad~$F\circ G$ on $\sTS$ is the identity
  and the adjunction restricts to an equivalence of categories between the full
  subcategory of~$\nsHDA{1}$ whose objects are strongly labeled.
\end{theorem}
\begin{proof}
  Suppose given a transition system~$T=(S,i,E,Tran)$ and a 1\nbd{}dimensional
  HDA~$C=(C,i)$. We construct a natural bijection between morphisms \hbox{$FC\to
    T$} in~$\sTS$ and morphisms~$C\to GT$ in~$\nsHDA{1}$. To every morphism
  $(\sigma,\tau):FC\to T$ of transition systems we associate the morphism of HDA
  $\varphi_{C,T}(\sigma,\tau):C\to GT$ defined as~$(\kappa,\lambda)$ where
  $\kappa_0=\sigma$ and $\kappa_1=\lambda=\tau$. Conversely, to every morphism
  $(\kappa,\lambda):C\to GT$ of HDA we associate the morphism
  $\psi_{C,T}(\kappa,\lambda):FC\to T$ of transition systems defined as
  $(\kappa_0,\lambda)$. These operations are mutually inverse and can easily be
  shown to be natural. The second part of the proposition can be checked
  directly.
\end{proof}

Now, recall that the category~$\TS$ can be defined as the Kleisli category
associated to the monad~$(-)_*$ on~$\sTS$. Similarly, the adjunction between
$\nsHDA{1}$ and~$\nHDA{1}$ given in Proposition~\ref{prop:psh-adj} induces a
monad~$\fpointed$ on~$\nsHDA{1}$ which ``replaces'' the underlying precubical
set of an HDA by the cubical set it generates.

\begin{theorem}
  \label{thm:ts-adj}
  The adjunction of Theorem~\eqref{thm:sts-adj} lifts to an adjunction
  between~$\TS$ and~$\nHDA{1}$, which induces an equivalence if we
  restrict~$\nHDA{1}$ to strongly labeled cubical sets.
\end{theorem}
\begin{proof}
  Consider a strict 1-dimensional HDA consisting of a precubical set~$C$ labeled
  by~$\ell$ into~$L$. Its image under the left adjoint~$F:\nsHDA{1}\to\nHDA{1}$
  to the forgetful functor $\nHDA{1}\to\nsHDA{1}$ is the 1-dimensional HDA whose
  underlying cubical set is~$D$ defined by~$D(0)=C(0)$, $D(1)=C(1)\uplus C(0)$
  with face maps being~$\partial_i^\alpha\uplus\id_{C(0)}:D(1)\to D(0)$ as face
  maps and the canonical injection~$\iota_0:D(0)\to D(1)$ as degeneracy maps,
  whose labeling is obtained by extending~$\ell_1:C(1)\to(\lcs L)_1$ to~$D(1)$
  by~$\ell_1(x)=(*)$ for $x\in C(0)\subseteq C(1)$. From this concrete
  description, it can easily be checked that $\fpointed\circ G=G\circ(-)_*$ and
  that the unit and the multiplication of $(-)_*$ are sent by~$G$ to the unit
  and multiplication of~$T$. Finally, we deduce that the adjunction of
  Theorem~\ref{thm:sts-adj} lifts to an adjunction between the Kleisli
  categories~$\TS$ and~$\nHDA{1}$ respectively associated to the monads $(-)_*$
  and $\fpointed$ using Proposition~\ref{prop:adj-lift}.
\end{proof}

\begin{remark}
  The fact that we have to restrict to a subcategory of $\nsHDA{1}$ in
  Theorem~\ref{thm:sts-adj} in order to obtain an equivalence of categories can
  be explained intuitively by remarking that in transition systems there is no
  distinction between events and labels: in particular, a transition system
  cannot contain two distinct transitions with the same event between the same
  source and the same target. For example, the following labeled (pre)cubical
  set cannot be represented in transition systems:
  \[
  \vxym{
    y\\
    x\ar@/_/_a[u]\ar@/^/^a[u]
  }
  \]
  More generally, in higher dimensions most models do not have the possibility
  to ``count'' the number of commutations between events: usually, two
  transitions either commute or not. This contrasts with cubical sets where a
  tile
  \[
  \svxym{
    &x_3&\\
    x_1\ar[ur]^{y_3}&&\ar[ul]_{y_4}x_2\\
    &x_0\ar[ul]^{y_1}\ar[ur]_{y_2}&
  }
  \]
  can be filled with many 2-cells. This explains why in the following most of
  the nice adjunctions will be obtained by restricting cubical sets to strongly
  labeled ones.
\end{remark}

By Proposition~\ref{prop:psh-adj}, the truncation functor~$\HDA\to\nHDA{1}$
admits a right adjoint. By composing this adjunction with the one of previous
theorem, we obtain an adjunction between~$\TS$ and~$\HDA$.

\begin{remark}
  The HDA associated by the right adjoint to a transition system
  \hbox{$T=(S,i,E,Tran)$} can be described in a more direct way using
  Proposition~\ref{prop:cts-hda} as generated by the cubical transition system
  \hbox{$(S,i,E,\ell,E\uplus\{*\},t)$} where \hbox{$\ell:E\to E\uplus\{*\}$} is
  the canonical injection and $t(x,l)=y$ if $l$ is reduced to an event~$e$ and
  $(x,e,y)\in Tran$, see Section~\ref{sec:cts} for details.
\end{remark}

\subsection{Asynchronous automata and HDA}
\label{sec:aa-cs}
The adjunction given in previous section, can be extended to an adjunction
between the category of strict asynchronous automata~$\sACR$ and the category of
strict 2-dimensional HDA~$\nsHDA{2}$.

To any strict 2-dimensional HDA~$C$, the left adjoint~$F:\nsHDA{2}\to\sACR$
associates the asynchronous automaton whose underlying transition system is
induced by the underlying 1-dimensional HDA of~$C$ and such that
$a_1\indep{s}a_2$ when there exists transitions $(s,a_1,s_1)$, $(s,a_2,s_2)$,
$(s_1,a_2,r)$ and $(s_2,a_1,r)$ and a $2$-cell~$y$ such that
$\partial_0^-(y)=(s,a_1,s_1)$, $\partial_0^+(y)=(s_2,a_1,r)$,
$\partial_1^-(y)=(s,a_2,s_2)$ and $\partial_1^+(y)=(s_1,a_2,r)$:
\begin{equation}
  \label{eq:aa-tile}
  \svxym{
    &r&\\
    s_1\ar[ur]^{a_2}&y&\ar[ul]_{a_1}s_2\\
    &\ar[ul]^{a_1}s\ar[ur]_{a_2}&\\
  }
\end{equation}
The functor is defined in the obvious way on morphisms.

Conversely, an asynchronous automaton \hbox{$A=(S,i,E,Tran,I)$} is sent by the
right adjoint $G:\sACR\to\nsHDA{2}$ to a strict $2$\nbd{}dimensional HDA~$C$,
whose underlying $1$-dimensional HDA is induced by the underlying transition
system of~$A$. The $2$-cells are \hbox{$C(2)= I$}, where~$I$ is seen as a subset
of~$E\times S\times E$. Given a pair of events~$a_1$ and $a_2$ related by~$I_s$
for some state~$s$, there exist transitions~$(s,a_1,s_1)$, $(s,a_2,s_2)$,
$(s_1,a_2,r)$ and $(s_2,a_1,r)$ and these are uniquely defined by the second
property of Definition~\ref{def:acr} as in~\eqref{eq:aa-tile},
face maps are defined on elements $y=a_1,s,a_2$ of~$I$ by
\[
\partial_0^-(y)=(s,a_1,s_1)
\quad
\partial_0^+(y)=(s_2,a_1,r)
\quad
\partial_1^-(y)=(s,a_2,s_2)
\quad
\partial_1^+(y)=(s_1,a_2,r)
\]
and the labeling function is defined by~$\ell(a_1,s,a_2)=(a_1,a_2)$. The
requirement that~$I$ is symmetric induces the symmetry of the HDA. The functor
is defined in the obvious way on morphisms.


\begin{theorem}
  These functors form an adjunction between~$\sACR$ and~$\nsHDA{2}$. The induced
  comonad on~$\sACR$ is the identity and the adjunction induces an equivalence
  of categories if we restrict~$\nsHDA{2}$ to the full subcategory of strongly
  labeled HDA. Moreover, this adjunction lifts to an adjunction between~$\ACR$
  and~$\nHDA{2}$ with similar properties.
\end{theorem}

\noindent
By composing with an adjunction given by Proposition~\ref{prop:psh-adj}, this
induces an adjunction between~$\ACR$ and~$\HDA$.

\subsection{Event structures and HDA}
We construct here an adjunction between~$\sLES$ and~$\sHDA$. This adjunction
reformulates in the framework of HDA some well-known relations between event
structures and transition systems with
independence~\cite{VSMNGWRelationships}. The study of relations between the two
models was initiated in~\cite{VSGLCHigher} and a similar connection is described
in~\cite{staton2010expressivity}.

A \emph{configuration} of an event structure~$(E,\leq,\incompat)$ is a finite
downward closed subset of compatible events in~$E$. An event~$e$ is
\emph{enabled} at a configuration~$x$ if~$e\not\in x$ and $x\uplus\{e\}$ is a
configuration. A functor~$F:\sLES\to\nsHDA{2}$ can be defined as follows. To any
labeled event structure~$(E,\leq,\incompat,\ell,L)$, it associates the
2-dimensional HDA~$C$ labeled by~$L$ whose $0$-cells are the configurations of
the event structure with the empty configuration as initial state, $1$-cells are
the pairs~$(x,e)$ where~$x$ is a configuration and~$e$ is an event enabled
at~$x$, and 2-cells are the pairs~$(x,e_1,e_2)$ where~$x$ is a configuration
and~$e_1,e_2$ are both enabled at~$x$ and such that~$e_2$ is enabled at
$x\uplus\{e_1\}$ and $e_1$ is enabled at $x\uplus\{e_2\}$, graphically:
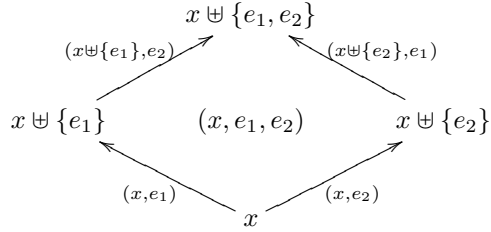
\[
\vxym{
  &x\uplus\{e_1,e_2\}&\\
  x\uplus\{e_1\}\ar[ur]^-{(x\uplus\{e_1\},e_2)}&(x,e_1,e_2)&x\uplus\{e_2\}\ar[ul]_-{(x\uplus\{e_2\},e_1)}\\
  &\ar[ul]^-{(x,e_1)}x\ar[ur]_-{(x,e_2)}&\\
}
\]
Notice for every 2-cell $(x,e_1,e_2)$, $(x,e_2,e_1)$ is also a 2-cell thus
inducing a symmetry on the precubical set. The functor is defined in the obvious
way on morphisms.

\begin{example}
  Consider the event structure $(E,\leq,\incompat,\ell,L)$, with
  $E=\{e_1,e_2,e_3\}$, with $e_1\leq e_2$ and~$e_1\leq e_3$, labeled
  in $L=\{a,b,c\}$ by $\ell(e_1)=a$, $\ell(e_2)=b$ and~$\ell(e_3)=c$. This event
  structure is represented on the left and induces the two HDA on the right
  when~$b$ is respectively incompatible and compatible with~$c$
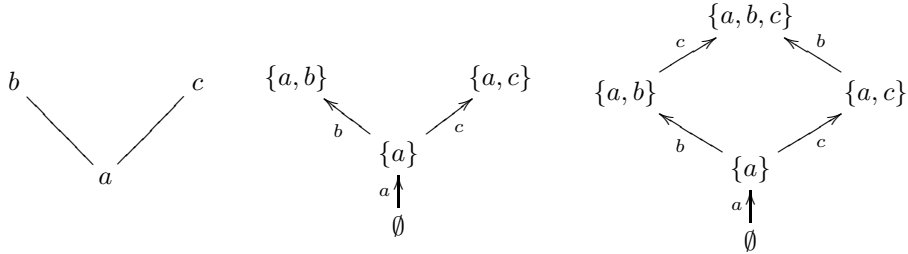
  \[
  \vxym{
    b&&c\\
    &\ar@{-}[ul]a\ar@{-}[ur]&\\
  }
  \hspace{4ex}
  \svxym{
    \\
    \{a,b\}&&\{a,c\}\\
    &\{a\}\ar[ul]^{b}\ar[ur]_c&\\
    &\emptyset\ar[u]^{a}\\
  }
  \hspace{4ex}
  \svxym{
    &\{a,b,c\}&\\
    \{a,b\}\ar[ur]^c&&\{a,c\}\ar[ul]_b\\
    &\{a\}\ar[ul]^{b}\ar[ur]_c&\\
    &\emptyset\ar[u]^{a}\\
  }
  \]
  (for simplicity we simply write~$e$ for a 1-cell~$(x,e)$ since~$x$ can be
  determined as the source of the cell). The square on the right diagram is
  filled with two 2-cells: $(\{a\},b,c)$ and~$(\{a\},c,b)$.
\end{example}

Conversely, we define a functor~$G:\nsHDA{2}\to\sLES$. The intuition is that
given an HDA~$C$, the elements of $G(C)$ should be the events of~$C$ in the
sense of Section~\ref{sec:cset}. However, event structures cannot express loops,
which should therefore be
unfolded~\cite{winskel-nielsen:models-concur,van2006expressiveness,fahrenberg2005higher}. For
example, an HDA of the form
\begin{equation}
  \label{eq:ex-loop}
  \vxym{
    x\ar@(ul,ur)^a
  }
\end{equation}
with only one 0-cell and one looping 1-cell should have as image an event
structure with a countable totally ordered set of events. A 2-dimensional HDA is
\emph{unfolded} when it is
\begin{itemize}
\item \emph{reachable}: every 0-cell~$x$ is reachable,
\item \emph{acyclic}: any path~$s:x\transitionpath{}x$ with the same source and
  target is empty,
\item \emph{unshared}: any two parallel paths~$s,t:x\transitionpath{}x'$ are
  homotopic.
\end{itemize}
This reformulates the notion of occurrence transition system with
independence. To any 2-dimensional HDA~$C$ with $i$ as initial state
and~$\ell:C\to\lcs L$ as labeling function, one can associate an unfolded
2\nbd{}dimensional HDA~$U(C)$ whose
\begin{itemize}
\item 0-cells are the paths $s:i\transitionpath{}x$ of~$C$ modulo homotopy,
\item 1-cells are the pairs $(s,m)$ where~$s:i\transitionpath{}x$ is a path
  and~$m$ is a 1-cell such that~$\partial_0^-(m)=x$, with $\partial_0^-(s,m)=s$
  and $\partial_0^+(s,m)=s\cdot m$ as source and target,
\item 2-cells are pairs~$(s,z)$ where~$s:i\transitionpath{}x$ is a path and $z$
  is a 2-cell such that~$\partial_0^-\partial_0^-(z)=x$, with its faces defined
  by $\partial_i^-(s,z)=(s,\partial_i^-(z))$ and
  $\partial_i^+(s,z)=(s\cdot\partial_{1-i}^-(z),\partial_i^+(z))$,
\item the labeling function labels a 1-cell $(s,m)$ by $\ell(m)$ and a 2-cell
  $(s,z)$ by $\ell(z)$.
\end{itemize}
This operation can easily be extended into a comonad on the
category~$\nsHDA{2}$. For example, the image of the HDA~\eqref{eq:ex-loop} is
\[
\vxym{
  x_0\ar[r]^{a}&x_1\ar[r]^a&x_2\ar[r]^a&\ldots
}
\]
Now, to every unfolded 2-dimensional HDA~$C$, one can associate a labeled event
structure~$V(C)=(E,\leq,\#,\ell,L)$ such that~$E$ is the set of events of~$C$ in
the sense of Section~\ref{sec:cset}. We say that an event~$e$ \emph{occurs} in a
path~$s$ when~$s$ contains a 1-cell~$m$ such that~$m\in e$. Two events~$e$
and~$e'$ are such that~$e\leq e'$ when for every path~$s\cdot
n:i\transitionpath{}x$ with~$n\in e'$ the event~$e$ occurs in~$s$. Two
events~$e$ and~$e'$ are such that~$e\#e'$ when there is no
path~$s:i\transitionpath{}x$ such that both~$e$ and~$e'$ occur in~$s$. The
labeling function is the labeling function of~$C$ (recall that we have shown in
Section~\ref{sec:lcs} that every labeled cubical set induces a labeling function
on its events). The operation~$V$ is easily extended as a functor~$V$ from the
category of unfolded 2-dimensional HDA to the category of labeled event
structures. Finally, we define the functor~$G:\nsHDA{2}\to\sLES$ as the
composite~$G=V\circ U$.


\begin{theorem}
  The composite functor $G\circ F$ is isomorphic to the identity functor
  on~$\sLES$. Thus~$\sLES$ embeds fully and faithfully into~$\nsHDA{2}$.
\end{theorem}
\begin{proof}
  The adjunction between labeled event structure and transition systems with
  independence described~\cite{VSMNGWRelationships,mimram:phd} can be
  straightforwardly adapted to asynchronous transition systems and one obtains
  the result by composing with the adjunction described in previous section.
\end{proof}

Notice that we did not claim that~$F$ and~$G$ are part of an adjunction, because
it is not the case. Namely, consider the effect of the endofunctor \hbox{$F\circ
  G:\nsHDA{2}\to\nsHDA{2}$}: we have pictured some HDA (on the left) together
with their image under~$F\circ G$ (on the right):
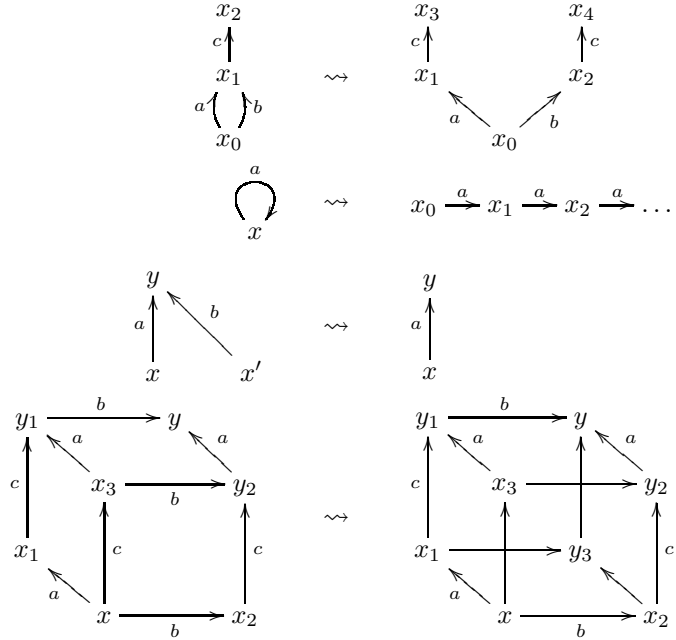
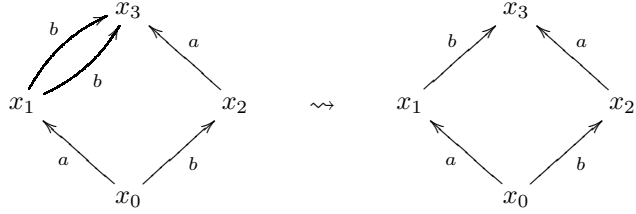
\begin{center}
  \begin{longtable}{r@{\qquad$\rightsquigarrow$\qquad}l}
    $
    \svxym{
      x_2\\
      x_1\ar[u]^c\\
      x_0\ar@/^/[u]^a\ar@/_/[u]_b
    }
    $
    &
    $
    \svxym{
      x_3&&x_4\\
      x_1\ar[u]^c&&x_2\ar[u]_c\\
      &x_0\ar[ul]^a\ar[ur]_b&\\
    }
    $
    \\[4ex]
    $
    \svxym{
      x\ar@(ul,ur)^a
    }
    $
    &
    $
    \svxym{
      x_0\ar[r]^a&x_1\ar[r]^a&x_2\ar[r]^a&\ldots
    }
    $
    \\[4ex]
    $
    \vxym{
      y\\
      x\ar[u]^a&x'\ar[ul]_b
    }
    $
    &
    $
    \vxym{
      y\\
      x\ar[u]^a
    }
    $
    \\[4ex]
    $
    \svxym{
      y_1\ar[rr]^b&&y\\
      &x_3\ar[ul]_a\ar[rr]_b&&y_2\ar[ul]_a\\
      x_1\ar[uu]^c&\\
      &\ar[ul]^{a}x\ar[uu]_c\ar[rr]_{b}&&x_2\ar[uu]_c\\
    }
    $
    &
    $
    \svxym{
      y_1\ar[rr]^b&&y\\
      &x_3\ar[ul]_a\ar[rr]&&y_2\ar[ul]_a\\
      x_1\ar[uu]^c\ar[rr]&&y_3\ar[uu]\\
      &\ar[ul]^{a}x\ar[uu]\ar[rr]_{b}&&x_2\ar[ul]\ar[uu]_c\\
    }
    $
    \\[10ex]
    $
    \vxym{
      &x_3&\\
      x_1\ar@/^/[ur]^{b}\ar@/_/[ur]_{b}&&\ar[ul]_{a}x_2\\
      &x_0\ar[ul]^{a}\ar[ur]_{b}&
    }
    $
    &
    $
    \vxym{
      &x_3&\\
      x_1\ar[ur]^b&&\ar[ul]_{a}x_2\\
      &x_0\ar[ul]^{a}\ar[ur]_{b}&
    }
    $
  \end{longtable}
\end{center}
In the third example, $x$ is the initial position and in the last two examples
all the squares for which it makes sense are filled with 2-cells. These examples
are representative of various kinds of behaviors that can happen:
\begin{itemize}
\item the first two examples show that ``shared transitions'' are ``unshared'',
  and in particular loops are unrolled;
\item the third example shows that the unreachable 0-cells of the HDA are
  removed,
\item the fourth example shows that if the HDA contains half of a cube then the
  other half of the cube is created, completing the cube -- this is related to
  the \emph{cube axiom} which is often used to characterize asynchronous
  transition systems generated by an event structure~\cite{VSMNGWRelationships};
\item the last example shows that HDA are made strongly labeled.
\end{itemize}
Notice, if we write~$C$ for the HDA in the left of examples, in the first three
examples there is a natural arrow~$TC\to C$ (but not in the other direction),
whereas in the last two examples there is a natural arrow $C\to TC$ (but not in
the other direction). So there is no hope that~$T$ would be either a monad or a
comonad, and thus that~$F$ and~$G$ either form an adjunction in either
direction.

It can however be shown that~$G$ is right adjoint to~$F$ if we
restrict~$\nsHDA{2}$ to the full subcategory whose objects are strongly labeled
and satisfy the \emph{cube axioms} (which state that if an asynchronous
transition system contains half of a cube as in fourth example then it also
contains the other half of the cube, as well as two other variants of this
property).
As previously, this adjunction can be extended to the non-strict variants of the
models, as well as the whole category~$\HDA$. This adjunction can also be
extended to an adjunction between general event structures (in which conflict is
not necessarily a binary relation) and HDA.


\subsection{Petri nets and HDA}
\label{sec:pnet-hda}
This section constitutes perhaps the most novel part of the paper. We extend
here previously constructed adjunctions between 1-bounded Petri Nets and
asynchronous transition
systems~\cite{winskel-nielsen:models-concur,MDRMSPetri,mukund:pn-sts,vanglabbeek:pn-cs-hda}
to an adjunction between general Petri Nets and HDA. For similar reasons as
previously, one needs to restrict to strongly labeled HDA in order to obtain a
well-defined adjunction. We thus implicitly only consider strongly labeled HDA
in the following.

\paragraph{Cubical transition systems.}
\label{sec:cts}
We introduce here a general methodology for associating a symmetric precubical
set to a model for concurrent processes, that we will use in order to associate
a strict HDA to a Petri net. Since monoidal functors preserve the unit of
monoidal categories, all cubical sets generated by cubical objects in~$\Set$
(\ie by the functor~$\lcs$ introduced in Section~\ref{sec:lcs}) contain only one
$0$\nbd{}cell. Cubical sets with multiple $0$-cells can be generated by
\emph{actions} of the labeling cubical set on the $0$-cells, formalized as
follows, in the same way that a transition system can be seen as an action of
the free monoid on labels over the states. The resulting notion of \emph{cubical
  transition system} (or \emph{CTS}) generalizes to the setting of cubical set
the notion of \emph{step transition system}~\cite{mukund:pn-sts} which is a
variant of transition systems in which multiple events can occur simultaneously.

\begin{definition}
  A \emph{cubical transition system}~$(S,i,E,t,\ell,L)$ consists of
  \begin{itemize}
  \item a set~$S$ of \emph{states},
  \item a state~$i\in S$ called the \emph{initial state},
  \item a set~$E$ of \emph{events},
  \item a \emph{transition function} which is a partial function~$t:S\times\slcs
    E\to S$,
  \item a set~$L$ of \emph{labels},
  \item a \emph{labeling function} $\ell:E\to L$,
  \end{itemize}
  such that for every state~$x$ and every $n$-cell~$l$ of~$\slcs E$ for
  which~$t(x,l)$ is defined,
  \begin{enumerate}
  \item if~$l=l_1\cdot l_2$ for some cells~$l_1$ and~$l_2$ then $t(x,l_1)$ and
    $t(t(x,l_1),l_2)$ are both defined and we have
    \hbox{$t(x,l)=t(t(x,l_1),l_2)$},
  \item $t(x,())$ is defined and equal to~$x$ (where $()$ denotes the 0-cell of
    $\slcs E$),
  \item for every symmetry~$\sigma:n\to n$, $t(x,\slcs E(\sigma)(l))$ is defined
    and equal to~$t(x,l)$.
  \end{enumerate}
\end{definition}

Cubical transition systems are thus generalized transition systems, which modify
state upon incoming events. These differ from traditional transition systems in
that they may accept a transition under $n$ events~$e_1,\ldots,e_n$, specified
by a transition under the word~$e_1\cdots e_n\in\slcs E$. With this
understanding in mind, the axioms have simple interpretations: for example the
first one states that the state reached under two simultaneous events~$e_1$
and~$e_2$ is the same as the state reached under~$e_1$ followed by~$e_2$.

An $n$-cell~$l$ of~$\slcs E$ is \emph{enabled} at a position~$x$ if~$t(x,l)$ is
defined.
Every such CTS defines a strict HDA~$C$ labeled by~$L$ whose $n$-cells are pairs
$(x,l)$ where~$x$ is a state and~$l$ is an $n$-cell of~$\slcs E$ which is
enabled at~$x$. Source and target functions are defined
by~$\partial_i^-(x,l)=(x,\partial_i^-(l))$ and
\hbox{$\partial_i^+(x,l)=t(t(x,e_i),\partial_i^+(l))$} where~$e_i$ is the $i$-th
element of~$l$
and symmetries by $\sigma(x,l)=(x,\slcs E(\sigma)(l))$. The labeling function
is~$\slcs\ell$ and the initial state is~$i$.

A \emph{morphism} $(\sigma,\tau,\lambda):(S_1,i_1,E_1,\ell_1,L_1,t_1) \to
(S_2,i_2,E_2,\ell_2,L_2,t_2)$ between two CTS consists of
\begin{itemize}
\item a function~$\sigma:S_1\to S_2$,
\item a function~$\tau:E_1\to E_2$,
\item a function~$\lambda:L_1\to L_2$,
\end{itemize}
such that~$i_2=\sigma(i_1)$, $\ell_2\circ\tau=\lambda\circ\ell_1$, and for every
state~$x\in S_1$ and cell~$l$ of~$\slcs E_1$,
$t_2(\sigma(x),\slcs\tau(l))=\sigma\circ t_1(x,l)$.
Every such morphism induces a morphism \hbox{$(\kappa,\lambda):C_1\to C_2$}
between the corresponding HDA~$C_1$ and~$C_2$ defined on $n$\nbd{}cells~$(x,l)$
of~$C_1$ by $\kappa_n(x,l)=(\sigma(x),\slcs\tau(l))$. We write~$\CTS$ for the
category thus defined.

\begin{proposition}
  \label{prop:cts-hda}
  The functor~$\CTS\to\sHDA$ defined above is well-defined.
\end{proposition}

\begin{remark}
  A variant of the notion of cubical transition system can easily be defined in
  order to generate symmetric cubical sets.
\end{remark}

\paragraph{From Petri nets to HDA.}
Suppose that we are given a labeled Petri net
$N=(P,M_0,E,\Pre,\Post,\ell,L)$. The~$\Pre{}$ and~$\Post{}$ operations can be
extended to the cells of~$\lcs E$ by \hbox{$\pre{()}=\pre{(*)}=0$},
$\pre{(l_1\cdot l_2)}=\pre{l_1}+\pre{l_2}$, $\post{()}=\post{(*)}=0$
and~$\post{(l_1\cdot l_2)}=\post{l_1}+\post{l_2}$. This enables us to see
elements of~$\lcs E$ as generalized events. We also generalize the notion of
transition and given two markings~$M_1$ and~$M_2$ and an event~$l\in\lcs E$, we
say that there is a transition \hbox{$M_1\transition{l}M_2$} whenever there
exists a marking~$M$ such that $M_1=M+\pre{l}$ and $M_2=M+\post{l}$. In this
case, the event~$l$ is said to be \emph{enabled} at the marking~$M_1$. The
marking~$M_2$ is sometimes denoted $M_1/l$. A marking~$M$ is \emph{reachable} if
there exists a transition~$l$ such that $M=M_0/l$ where~$M_0$ is the initial
marking of~$N$.

\begin{remark}
  \label{rem:pn-morph-direction}
  As in~\cite{winskel-nielsen:models-concur}, we have chosen to define morphisms
  in the opposite direction on places. With the adjunction with HDA in mind,
  this can be explained as follows. Morphisms of Petri nets should, just as
  morphisms of HDA, preserve independence of events: if two events~$e$ and~$e'$
  of a net~$N$ are independent and~$(\varphi,\psi):N\to N'$ is a morphism of
  nets, then their images~$\psi(e)$ and~$\psi(e')$ should also be
  independent. By contraposition, this means that if both events~$\psi(e)$
  and~$\psi(e')$ depend on a common place~$p$, then the events~$e$ and~$e'$
  should depend on a corresponding common place~$\psi^{-1}(p)$.
\end{remark}

Every labeled Petri net~$N$ induces a CTS~$(S,i,E,t,\ell,L)$ whose states~$S$
are the reachable markings of the net, with the initial marking $M_0$ as initial
state, events~$E$ are the events of the net, transition function $t(M,l)$ is
defined if and only if $l$ is enabled at~$M$ and in this case
\hbox{$t(M,l)=M/l$}, with the set~$L$ as set of labels and~$\ell:E\to L$ as
labeling function.

It is routine to verify that this actually defines a CTS and thus a strict
HDA. The $n$-cells of~$\hda(N)$ consisting of a marking~$M$ of the net and a
list~$l$ of events which is enabled at~$M$. Moreover, any morphism
$(\varphi,\psi):N\to N'$ between labeled Petri nets induces a morphism
$(\sigma,\tau,\lambda)$ between the corresponding CTS defined by
\hbox{$\sigma(M)=M\circ\varphi$} for any reachable marking~$M$ of~$N$,
$\tau=\psi$, and~$\lambda=\psi$. We denote by \hbox{$\hda:\sLPNet\to\sHDA$} the
functor thus defined.


\paragraph{From HDA to Petri nets.}
We first introduce the notion of region of an HDA, which should be thought as a
way of associating a number of tokens to each 0-cell of the HDA and a pre- and
postcondition to every transition of the HDA, in a coherent way. A
pre-region~$R$ of a precubical set~$C$ is a sequence~$(R_i)_{i\in\N}$ of
functions~$R_i:C(i)\to\N\times\N$ such that
\begin{itemize}
\item for every~$x\in C(0)$, $R_0(x)=(0,0)$
\item for every~$x\in C(i+1)$ and $\alpha_k\in\{-,+\}$,
  \[
  R_{i+1}(x)\qeq\sum_{k=0}^iR_1(\partial_{\lnot k}^{\alpha_k}(x))
  \]
  where the sum is computed coordinate by coordinate on pairs of integers.
\end{itemize}
Notice that, by the second property, a region is uniquely determined by the
image of $1$-dimensional cells in~$x\in C(1)$.
We sometimes omit the index~$i$ since it is determined by the dimension of the
cell in argument and respectively write~$R'(x)$ and~$R''(x)$ for the first and
second components of~$R(x)$, where~$x$ is a cell of~$C$. It can be remarked that
two 1-cells which are part of the same event necessarily have the same image
under a pre-region; a pre-region~$R$ thus induces a function from the events
of~$C$ to $\N\times\N$, that we still write~$R$.
A \emph{region} of a precubical set consists of a pre-region~$R$ together with a
function \hbox{$S:C(0)\to\N$} such that for every $i$-cell $y\in C(i)$ whose
0-source is~$x$ and 0-target is~$x'$, there exists an integer~$n$ such that
\hbox{$(S(x),S(x'))=(n+R'(y),n+R''(y))$}.

\begin{example}
  Consider the following precubical set
  \[
  \vxym{
    &x_3&\\
    x_1\ar[ur]^{y_1}&z&x_2\ar[ul]_{y_3}\\
    &\ar[ul]^{y_0}x_0\ar[ur]_{y_2}\ar[rr]^{y_4}&&x_4\\
  }
  \]
  A region~$(R,S)$ for this cubical set is for example
  \[
  R(y_0)=(2,1)
  \quad
  R(y_1)=(3,1)
  \quad
  R(y_2)=(3,1)
  \quad
  R(y_3)=(2,1)
  \quad
  R(y_4)=(0,2)
  \]
  and
  \[
  R(z)=(5,2)
  \quad
  S(x_0)=6
  \quad
  S(x_1)=5
  \quad
  S(x_2)=4
  \quad
  S(x_3)=3
  \quad
  S(x_4)=8
  \]
  Graphically,
  \[
  \vxym{
    &3&\\
    5\ar[ur]^{(3,1)}&(5,2)&4\ar[ul]_{(2,1)}\\
    &\ar[ul]^{(2,1)}6\ar[ur]_{(3,1)}\ar[rr]^{(0,2)}&&8\\
  }
  \]
\end{example}

To every strict HDA~$C$, we associate a labeled Petri net~$\pn(C)$ whose
\begin{itemize}
\item places are the regions of~$C$,
\item events are the events of~$C$, labeled as in~$C$,
\item pre and post functions are given on any event~$e$ and any place~$(R,S)$ by
  \hbox{$\pre{e}(R,S)=R'(e)$} and \hbox{$\post{e}(R,S)=R''(e)$},
\item initial marking~$M_0$ is~$M_0(R,S)=S(x_0)$, where~$x_0$ is the initial
  state of~$C$.
\end{itemize}
Suppose that \hbox{$(\kappa,\lambda):C\to D$} is a morphism of HDA. We define a
morphism of labeled Petri nets \hbox{$\pn(\kappa,\lambda):\pn(C)\to\pn(D)$} as
follows: $\pn(\kappa,\lambda)=(\varphi,\psi,\lambda)$, where
\begin{itemize}
\item $\varphi$ maps every region~$(R,S)$ of~$D$ to the region
  \hbox{$\varphi(R,S)=(R\circ\kappa,S\circ\kappa_0)$}, where~$R\circ\kappa$
  denotes the pre-region~$(R_i\circ\kappa_i)_{i\in\N}$,
\item $\psi$ is the map induced on events by~$\kappa_1$ (two 1-cells which are
  part of the same event are sent to 1-cells which are part of the same event
  by~$\kappa_1$).
\end{itemize}
This thus defines a functor \hbox{$\pn:\sHDA\to\sLPNet$}.

\subsubsection{The adjunction.}
Suppose that we are given an HDA~$C$ labeled by~$\ell$ into~$L$, and a labeled
net $N=(P,M_0,E,\Pre{},\Post{},m,M)$. We want to exhibit a bijection between
morphisms \hbox{$\pn(C)\to N$} in~$\sLPNet$ and morphisms~$C\to\hda(N)$
in~$\sHDA$.

To any morphism~$(\varphi,\psi,\lambda):\pn(C)\to N$ of labeled Petri nets, we
associate a morphism \hbox{$(\kappa,\lambda):C\to\hda(N)$} of HDA defined as
follows. Given an $n$\nbd{}cell~$x$ of~$C$, $\kappa_n(x)$ should be an $n$-cell
of $\hda(N)$, that is a pair $(M_{\kappa_n(x)},l_{\kappa_n(x)})$ where
$M_{\kappa_n(x)}$ is a marking of~$N$ and $l_{\kappa_n(x)}$ is a list of events
of~$N$ which is enabled at~$M_{\kappa_n(x)}$. These are defined for every place
$p$ of~$N$ by \hbox{$M_{\kappa_n(x)}(p)=S_{\varphi(p)}(y)$}, where $y$ is the
0-source of~$x$, and $l_{\kappa_n(x)}=\lcs\psi(\overline{\partial_{\lnot
    0}^-(x)} \cdots \overline{\partial_{\lnot(n-1)}^-(x)})$ where~$\overline{y}$
denotes the event associated to a 1-cell~$x$.


Conversely, to any morphism~$(\kappa,\lambda):C\to\hda(N)$ of HDA, we associate
a morphism of labeled Petri nets \hbox{$(\varphi,\psi,\lambda):\pn(C)\to N$}
defined as follows. Given an $n$-cell~$x$, $\kappa_n(x)$ is an $n$-cell of
$\hda(N)$, that is a pair $(M_{\kappa_n(x)},l_{\kappa_n(x)})$ as above. For
every place~$p$, $\varphi(p)$ is the region~$(R_{\varphi(p)},S_{\varphi(p)})$
of~$C$ which is defined on 0-cells~$x$ by~$S_{\varphi(p)}(x)=M_{\kappa_n(x)}(p)$
and on $n$-cells~$x$ by
$R_{\varphi(p)}=(\pre{l_{\kappa_n(x)}},\post{l_{\kappa_n(x)}})$. Given a
1-cell~$x$, its image under~$\kappa_1(x)$ is a pair
$(M_{\kappa_1(x)},l_{\kappa_1(x)})$ where~$l_{\kappa_1(x)}$ is reduced to one
1-cell~$y$. It is immediate to check that for any other 1\nbd{}cell~$x'$ such
that $x\approx x'$, we have that $l_{\kappa_1(x)}\approx l_{\kappa_1(x')}$: it
thus makes sense to extend~$x\mapsto l_{\kappa_1(x)}$ into a function which to
an event~$e$ of~$C$ associates an event~$l_{\kappa_1(e)}$. Given an event~$e$
of~$C$, we define~$\psi(e)=l_{\kappa_1(e)}$.

It can be shown that these transformations are well defined, are natural in~$C$
and~$N$, and are mutually inverse. Therefore,

\begin{theorem}
  The functor \hbox{$\hda:\sLPNet\to\sHDA$} is right adjoint to the functor
  \hbox{$\pn:\sHDA\to\sLPNet$}.
\end{theorem}
\begin{proof}
  It is routine to check that the transformations given above are well-defined
  and natural in~$C$ and~$N$. We now show that they are mutually inverse.

  Suppose that~$(\varphi,\psi,\lambda):\pn(C)\to N$ is a morphism of Petri nets
  and consider the associated morphisms
  \[
  (\kappa,\lambda):C\to\hda(N)
  \qqtand
  (\varphi',\psi',\lambda):\pn(C)\to N
  \]
  obtained by successively applying the two transformations above. For any
  place~$p$ of~$N$, $\varphi'(p)$ is a place of~$\pn(C)$, that is a region
  $(R_{\varphi'(p)},S_{\varphi'(p)})$ of~$C$. By definition of the
  transformations, we have that for every $0$-cell~$x$ of~$C$,
  $S_{\varphi'(p)}(x)=M_{\kappa_n(x)}(p)=S_{\varphi(p)}(x)$ and for every
  $n$-cell~$y$ of~$C$, the first component of~$R_{\varphi'(p)}(x)$ is
  \[
  \pre{l_{\kappa_n(x)}}(p)=
  \sum_{i=0}^{n-1}\Pre\circ\psi(\overline{\partial_i^-(x)})(p)=
  \sum_{i=0}^{n-1}\Pre(\overline{\partial_i^-(x)})(\varphi(p))= R_{\varphi(p)}'
  \]
  and similarly $\post{l_{\kappa_n(x)}}(p)= R_{\varphi(p)}''$, thus
  $R_{\varphi'(p)}=R_{\varphi(p)}$. Moreover, for every event~$e$ of~$C$,
  $\psi'(e)= \psi(e)$.

  Conversely, suppose that $(\kappa,\lambda):C\to\hda(N)$ is a morphism of
  cubical sets and consider the associated morphisms
  \[
  (\varphi,\psi,\lambda):\pn(C)\to N
  \qqtand
  (\kappa',\lambda):C\to\hda(N)
  \]
  obtained by successively applying the two transformations above. For any
  $n$-cell~$x$ of~$C$, the $n$-cell $\kappa'_n(x)$ is an $n$-cell of~$\hda(N)$
  consisting of a pair $(M_{\kappa'_n(x)},l_{\kappa'_n(x)})$ as above. By
  definition of~$\hda(N)$, we have $M_{\kappa'_n(x)}=M_{\kappa'_n(y)}$,
  where~$y$ is the 0-source of~$x$. Moreover, for every place~$p$ of~$N$, we
  have $M_{\kappa'_n(y)}(p)= S_{\varphi(p)}(y)= M_{\kappa_n(y)}(p)$. And
  finally,
  \[
  l_{\kappa'_n(x)}= \lcs\psi(\overline{\partial_{\lnot 0}^-(x)} \cdots
  \overline{\partial_{\lnot(n-1)}^-(x)})= (\kappa_1(\overline{\partial_{\lnot
      0}^-(x)}) \cdots \kappa_1(\overline{\partial_{\lnot(n-1)}^-(x)}))=
  l_{\kappa_n(x)}
  \]
  which concludes the construction of the adjunction.
\end{proof}

\begin{example}
  If we restrict to 1-bounded nets, which are nets a place can contain either
  $0$ or $1$ token, we can recover the constructions
  of~\cite{winskel-nielsen:models-concur} for constructing an adjunction between
  asynchronous transition systems and nets. Since the net associated to an HDA
  by the functor hda is generally infinite, we will give an example in the case
  of 1-bounded nets. Consider the asynchronous automaton, depicted on the left
  of~\eqref{eq:ex-hda-pn}, with an empty independence relation.
  \begin{equation}
    \label{eq:ex-hda-pn}
    \vcenter
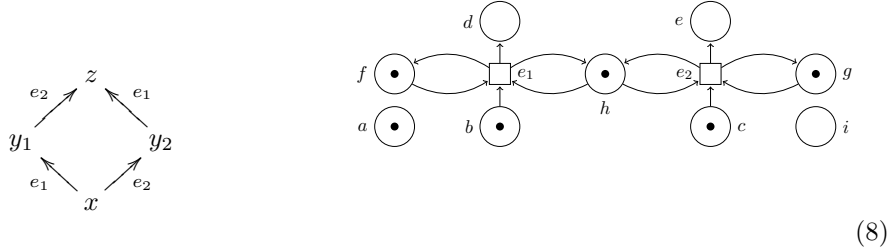
{
      \xymatrix@C=3ex@R=3ex{
        &z&\\
        y_1\ar[ur]^{e_2}&&\ar[ul]_{e_1}y_2\\
        &\ar[ul]^{e_1}x\ar[ur]_{e_2}&
      }
    }
    \qquad\qquad\qquad
    \scalebox{0.7}{
      \begin{tikzpicture}
        \node[place,tokens=1,label=left:$a$] (a) at (0,0) {};
        \node[place,tokens=1,label=left:$b$] (b) at (2,0) {};
        \node[place,tokens=1,label=right:$c$] (c) at (6,0) {};
        \node[place,tokens=0,label=left:$d$] (d) at (2,2) {};
        \node[place,tokens=0,label=left:$e$] (e) at (6,2) {};
        \node[place,tokens=1,label=left:$f$] (f) at (0,1) {};
        \node[place,tokens=1,label=right:$g$] (g) at (8,1) {};
        \node[place,tokens=1,label=below:$h$] (h) at (4,1) {};
        \node[place,tokens=0,label=right:$i$] (i) at (8,0) {};
        \node[transition,label=right:$e_1$] at (2,1) {}
        edge[pre] (b)
        edge[post] (d)
        edge[pre,bend left] (f)
        edge[post,bend right] (f)
        edge[pre,bend right] (h)
        edge[post,bend left] (h);
        \node[transition,label=left:$e_2$] at (6,1) {}
        edge[pre] (c)
        edge[post] (e)
        edge[pre,bend right] (g)
        edge[post,bend left] (g)
        edge[pre,bend left] (h)
        edge[post,bend right] (h);
      \end{tikzpicture}
    }
  \end{equation}
The associated 1-bounded Petri net is shown on the right. In this automaton the
place~$d$ corresponds to the region~$(R,S)$ such that~$R(e_1)=(1,0)$,
$R(e_2)=(0,0)$, $S(x)=S(y_2)=1$ and $S(y_1)=S(z)=0$. Graphically,
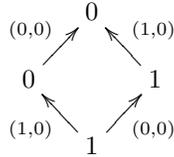
\[
\vcenter{
  \xymatrix@C=3ex@R=3ex{
    &0&\\
    0\ar[ur]^{(0,0)}&&\ar[ul]_{(1,0)}1\\
    &\ar[ul]^{(1,0)}1\ar[ur]_{(0,0)}&
  }
}
\]
Now, if we consider the same automaton with $e_1 \indep{x} e_2$, we obtain the
same Petri net with the place~$h$ removed. The general (\ie non-bounded) net
associated to an HDA is generally infinite (even for very simple examples) and
thus difficult to describe, which is why we did not provide an example in the
general case.
\end{example}

\noindent
This adjunction can easily be lifted into an adjunction between~$\LPNet$
and $\HDA$.

\section{Conclusion and future work}
\label{sec:conclusion}
In this paper, we have made completely formal the relation between HDA and
various classical models of concurrent computations: transition systems,
asynchronous automata, event structures and Petri nets. This is not only
interesting for comparison purposes, between different semantics of parallel
languages, but also, for practical reasons, which will be detailed in a
subsequent article.

Stubborn sets \cite{AVStubborn2}, sleep sets and persistent sets
\cite{PGPWUsing} are methods used for diminishing the complexity of
model-checking using transition systems. They are based on semantic observations
using Petri nets in the first case and Mazurkie\-wicz trace theory in the other
one. We believe that these are special forms of ``homotopy retracts'' when cast
(using the adjunctions we have hinted) in the category of higher-dimensional
transition systems. We shall make this statement more formal through these
adjunctions, which will allow for new state-space reduction methods.

Last but not least, in \cite{AJMNGWBisimulation} is defined an abstract notion
of bisimulation. Given a model for concurrency, i.e. a category of models $\bf
M$ and a ``path category'' (a subcategory of $\bf M$ which somehow represents
what should be thought of as being paths in the models), then we can define two
elements of $\bf M$ to be bisimilar if there exists a span of special morphisms
linking them. These special morphisms have a path-lifting property that, we
believe, would be in higher-dimensional transition systems a (geometric)
fibration property. We thus hope that homotopy invariants could be useful for
the study of a variety of bisimulation equivalences (some work has been done in
that direction in~\cite{VSGLCHigher,fahrenberg2005category}).

\bibliographystyle{plain}
\bibliography{biblio}

\begin{thebibliography}{10}

\bibitem{AASystemes}
A.~Arnold.
\newblock {\em Syst\`emes de transitions finis et s\'emantique des processus
  communicants}.
\newblock Masson, 1992.

\bibitem{MABCategories}
M.~A. Bednarczyk.
\newblock {\em Categories of asynchronous systems}.
\newblock PhD thesis, 1988.

\bibitem{borceux2001galois}
F.~Borceux and G.~Janelidze.
\newblock {\em {Galois theories}}.
\newblock Cambridge Univ Press, 2001.

\bibitem{RBPJHAlgebra}
R.~Brown and P.~J. Higgins.
\newblock On the algebra of cubes.
\newblock {\em JPAA}, (21):233--260, 1981.

\bibitem{VSGLCHigher}
G.L. Cattani and V.~Sassone.
\newblock Higher dimensional transition systems.
\newblock In {\em Eleventh Annual IEEE Symposium on Logic in Computer Science
  (LICS'96)}, pages 55--62, 1996.

\bibitem{MDRMSPetri}
M.~Droste and RM~Shortt.
\newblock {Petri nets and automata with concurrency relations—an adjunction}.
\newblock In {\em Sem. of Prog. Lang. and Model Theory}, pages 69--87, 1993.

\bibitem{fahrenberg2005category}
U.~Fahrenberg.
\newblock A category of higher-dimensional automata.
\newblock {\em Foundations of Software Science and Computational Structures},
  pages 187--201, 2005.

\bibitem{fahrenberg2005higher}
U.~Fahrenberg.
\newblock {\em Higher-Dimensional Automata from a Topological Viewpoint}.
\newblock PhD thesis, Aalborg University, 2005.

\bibitem{LFMSCS}
L.~Fajstrup.
\newblock Loops, ditopology, and deadlocks.
\newblock {\em Math. Struct. Comput. Sci.}, 2000.

\bibitem{fajstrup2005dipaths}
L.~Fajstrup.
\newblock Dipaths and dihomotopies in a cubical complex.
\newblock {\em Advances in Applied Mathematics}, 35(2):188--206, 2005.

\bibitem{LFEGMRDetecting}
L.~Fajstrup, E.~Goubault, and M.~Rau{\ss}en.
\newblock {Detecting deadlocks in concurrent systems}.
\newblock {\em CONCUR'98}, pages 332--347.

\bibitem{LFEGMRAlgebraic}
L.~Fajstrup, M.~Rau{\ss}en, and E.~Goubault.
\newblock {Algebraic topology and concurrency}.
\newblock {\em Theoretical Computer Science}, 357(1-3):241--278, 2006.

\bibitem{gaucher2010combinatorics}
P.~Gaucher.
\newblock Combinatorics of labelling in higher-dimensional automata.
\newblock {\em Theoretical Computer Science}, 411(11-13):1452--1483, 2010.

\bibitem{PGPWUsing}
P.~Godefroid and P.~Wolper.
\newblock Using partial orders for the efficient verification of deadlock
  freedom and safety properties.
\newblock volume 575, pages 417--428. LNCS, 1991.

\bibitem{EGGeometry}
E.~Goubault.
\newblock {\em The Geometry of Concurrency}.
\newblock PhD thesis, 1995.

\bibitem{EGCubical}
E.~Goubault.
\newblock Cubical sets are generalized transition systems.
\newblock Technical report, 2001.

\bibitem{goubault2001labelled}
E.~Goubault.
\newblock Labelled cubical sets and asynchronous transistion systems: an
  adjunction.
\newblock In {\em Presented at CMCIM'02}, volume~2, page 2002, 2001.

\bibitem{EGTPJHomology}
E.~Goubault and T.~P. Jensen.
\newblock Homology of higher-dimensional automata.
\newblock In {\em Proc. of CONCUR'92}, Stonybrook, New York, August 1992.
  Springer-Verlag.

\bibitem{Grandisbook}
M.~Grandis.
\newblock {\em Directed Algebraic Topology; models of non-reversible worlds}.
\newblock \!CUP, 2009.

\bibitem{grandis-mauri:cubical-sets}
M.~Grandis and L.~Mauri.
\newblock {Cubical sets and their site}.
\newblock {\em TAC}, 11(8):185--211, 2003.

\bibitem{haymansymmetry}
J.~Hayman and G.~Winskel.
\newblock Symmetry in petri nets.
\newblock {\em Perspectives in Concurrency Theory}, 2008.

\bibitem{AJMNGWBisimulation}
A.~Joyal, M.~Nielsen, and Winskel G.
\newblock Bisimulation and open maps.
\newblock In {\em LICS}, 1993.

\bibitem{mac1992sheaves}
S.~Mac~Lane and I.~Moerdijk.
\newblock {\em Sheaves in geometry and logic: A first introduction to topos
  theory}.
\newblock Springer, 1992.

\bibitem{maclane:cwm}
S.~MacLane.
\newblock {\em Categories for the {W}orking {M}athematician}, volume~5 of {\em
  Graduate Texts in Mathematics}.
\newblock Springer Verlag, 1971.

\bibitem{mimram:phd}
S.~Mimram.
\newblock {\em {S{\'e}mantique des jeux asynchrones et r{\'e}{\'e}criture
  2\nobreakdash-dimensionnelle}}.
\newblock PhD thesis, PPS, CNRS -- Universit{\'e} Paris Diderot, 2008.

\bibitem{mukund:pn-sts}
M.~Mukund.
\newblock {Petri nets and step transition systems}.
\newblock {\em International Journal of Foundations of Computer Science},
  3(4):443--478, 1992.

\bibitem{mulry:lifting-kleisli}
P.~Mulry.
\newblock {Lifting theorems for Kleisli categories}.
\newblock In {\em MFPS}, pages 304--319, 1994.

\bibitem{event1}
M.~Nielsen, G.~D. Plotkin, and G.~Winskel.
\newblock Petri nets, event structures and domains.
\newblock In {\em Semantics of Concurrent Computation}, pages 266--284, 1979.

\bibitem{carl}
C.~Petri.
\newblock Communication with automata, 1966.

\bibitem{VPModeling}
V.~Pratt.
\newblock Modeling concurrency with geometry.
\newblock In {\em Proc. of the 18th ACM Symposium on Principles of Programming
  Languages}. ACM Press, 1991.

\bibitem{MRMSCS}
M.~Raussen.
\newblock On the classification of dipaths in geometric models for concurrency.
\newblock {\em Mathematical Structures in Computer Science}, August 2000.

\bibitem{invPetri}
Sankaranarayanan S., H.~Sipma, and Z.~Manna.
\newblock Petri net analysis using invariant generation.
\newblock volume 2772 of {\em LNCS}, pages 682--701, 2003.

\bibitem{VSMNGWRelationships}
V.~Sassone, M.~Nielsen, and G.~Winskel.
\newblock Relationships between models of concurrency.
\newblock In {\em Proceedings of the Rex'93 school and symposium}, 1994.

\bibitem{JPSHomologie}
J.P. Serre.
\newblock {\em Homologie Singuli\`ere des Espaces Fibr\'es. Applications}.
\newblock PhD thesis, \'Ecole Normale Sup\'erieure, 1951.

\bibitem{MWSConcurrent}
M.W. Shields.
\newblock Concurrent machines.
\newblock {\em Computer Journal}, 28, 1985.

\bibitem{staton2010expressivity}
S.~Staton and G.~Winskel.
\newblock {On the expressivity of symmetry in event structures}.
\newblock In {\em Logic in Computer Science}, pages 392--401. Citeseer, 2010.

\bibitem{AVStubborn2}
A.~Valmari.
\newblock A stubborn attack on state explosion.
\newblock In {\em Proc. of CAV'90}. LNCS, 1990.

\bibitem{RVGBisimulation}
R.~van Glabbeek.
\newblock Bisimulation semantics for higher dimensional automata.
\newblock Technical report, Stanford University, 1991.

\bibitem{vanglabbeek:pn-cs-hda}
R.~van Glabbeek.
\newblock {Petri nets, configuration structures and higher dimensional
  automata}.
\newblock {\em Lecture notes in computer science}, pages 21--27, 1999.

\bibitem{van2006expressiveness}
R.~van Glabbeek.
\newblock On the expressiveness of higher dimensional automata.
\newblock {\em Theoretical computer science}, 356(3):265--290, 2006.

\bibitem{event2}
G.~Winskel.
\newblock Event structures.
\newblock In {\em Advances in Petri Nets}, pages 325--392, 1986.

\bibitem{winskel2007event}
G.~Winskel.
\newblock Event structures with symmetry.
\newblock {\em Electronic Notes in Theoretical Computer Science}, 172:611--652,
  2007.

\bibitem{winskel-nielsen:models-concur}
G.~Winskel and M.~Nielsen.
\newblock Models for concurrency.
\newblock In {\em Handbook of Logic in Computer Science}, volume~3, pages
  1--148. Oxford University Press, 1995.

\end{thebibliography}
\end{document}